\documentclass[aps,prd,preprint,superscriptaddress,showpacs,floatfix,
preprintnumbers,nofootinbib, a4paper]{revtex4-1}
\usepackage{color}
\usepackage{graphicx}
\usepackage{textcomp}
\usepackage{subfigure}
\usepackage{feynmp}
\usepackage{amsmath,amssymb,array}
\usepackage{enumerate}
\newcommand{\mathsym}[1]{{}} 
\def\lsim{\:\raisebox{-1.1ex}{$\stackrel{\textstyle<}{\sim}$}\:}

\newcommand{\ba}{\begin{array}} 
\newcommand{\ea}{\end{array}}
\newcommand{\be}{\begin{equation}}
\newcommand{\ee}{\end{equation}}
\newcommand{\beqa}{\begin{eqnarray}} 
\newcommand{\eeqa}{\end{eqnarray}}

\def\vev#1{\left\langle #1\right\rangle}

\def\3{$\bf{3}$}
\def\R3p{$\bf{3}^\bf{\prime}$}

\begin{document} 
\vspace*{1cm}
\title{Discrete symmetries for electroweak natural type-I seesaw mechanism} 
\bigskip 
\author{Pratik Chattopadhyay}
\email{pratikchattopadhyay@iisermohali.ac.in} 
\author{Ketan M. Patel}
\email{ketan@iisermohali.ac.in} 
\affiliation{Indian Institute of Science Education and Research Mohali, Knowledge City, Sector  81, S A S Nagar, Manauli 140306, India.}

\begin{abstract}
The naturalness of electroweak scale in the models of type-I seesaw mechanism with ${\cal O}(1)$ Yukawa couplings requires TeV scale masses for the fermion singlets. In this case, the tiny neutrino masses have to arise from the cancellations within the seesaw formula which are arranged by fine-tuned correlations between the Yukawa couplings and the masses of fermion singlets. We motivate such correlations through the framework of discrete symmetries. In the case of three Majorana fermion singlets, it is shown that the exact cancellation arranged by the discrete symmetries in seesaw formula necessarily leads to two mass degenerate fermion singlets. The remaining fermion singlet decouples completely from the standard model. We provide two candidate models based on the groups $A_4$ and $\Sigma(81)$ and discuss the generic perturbations to this approach which can lead to the viable neutrino masses.
\end{abstract} 

\maketitle

\section{Introduction} 
\label{intro}
The type-I seesaw mechanism is considered to be one of the  simplest and minimal extension of the Standard Model (SM) that can naturally generate small neutrino masses \cite{Minkowski:1977sc, Yanagida:1979as, Glashow:1979nm, Mohapatra:1979ia, GellMann:1980vs,Schechter:1980gr,Schechter:1981cv}. It requires existence of new fermions, often called as right handed (RH) neutrinos, which are singlets under the SM gauge symmetry. These fermions can have large Majorana mass and they couple to the SM only through their Yukawa interactions with the leptons and Higgs. If such couplings are taken to be of the same order of other Yukawa couplings in the SM, the singlet fermions are required to be as massive as $10^{9}$-$10^{15}$ GeV in order to comply with the neutrino mass scale that governs solar and atmospheric neutrino oscillations. This makes it almost impossible to verify the existence of RH neutrinos experimentally and in turn also the validity of type-I seesaw mechanism. The low scale versions of type-I seesaw mechanism have been also put forward in which the masses of fermion singlets are assumed to be at experimentally accessible scales \cite{delAguila:2005ssc,Bray:2005wv,Han:2006ip,delAguila:2006bda,Atwood:2007zza,Bray:2007ru,deAlmeida:2007gfv,delAguila:2007qnc,Bajc:2007zf}. For the recent study of the phenomenology of light fermion singlets, see \cite{Deppisch:2015qwa,Das:2017nvm} and references therein. In these versions, the smallness of neutrino mass is arranged by assuming either very small Yukawa couplings or cancellations in the seesaw mass formula \cite{Buchmuller:1991tu,Ingelman:1993ve,Heusch:1993qu} which arise due to some very particular choices of Yukawa couplings and masses of the fermion singlets. The tiny Yukawa couplings lead to very small mixing between the SM neutrinos and fermion singlets making their production suppressed in the direct search experiments. While the possibility of seesaw cancellations with ${\cal O}(1)$ Yukawa couplings and light fermion singlets seems quite promising from experimental point of view, it remains very highly fine-tuned if such cancellations are not motivated from some symmetry or dynamical mechanisms. In any case, demanding low-scale seesaw mechanism based on the grounds of only  experimental accessibility is not well motivated as it goes against the very basic idea of seesaw mechanism for which it was actually proposed, namely to naturally suppress neutrino masses through introducing a heavy scale in the theory.

A more profound constraint on type-I seesaw mechanism comes from the requirement of electroweak naturalness. The discovery of boson with mass 126 GeV which seems very much like the SM Higgs \cite{Aad:2012tfa,Chatrchyan:2012xdj,Khachatryan:2014jba} validates the idea of spontaneous breaking of electroweak symmetry through Brout–Englert–Higgs mechanism. The Higgs field $\phi$ has potential $V=-\mu^2 |\phi|^2+ \lambda |\phi|^4$ where $\mu$ is a dimensionful and $\lambda$ is a dimensionless parameter. They set vacuum expectation value (VEV) of Higgs field $v = \sqrt{\mu^2/\lambda} \equiv 246 $ GeV which is  determined from the masses of $W$ and $Z$ bosons. They also determine the mass of physical Higgs boson, namely $M_\phi = 2 \lambda v^2$. The measurement of Higgs mass therefore completely determines the Higgs potential and implies the value for renormalized dimensionfull parameter, $\mu = M_\phi/\sqrt{2} \approx 89 $ GeV. The hierarchy or electroweak naturalness problem refers to the higher order corrections to the $\mu^2$ parameter and concerns to the stability of this scale under such corrections. If SM is the only fundamental theory extendable to any arbitrary high scale then there is no naturalness problem. However in the presence of any new physics beyond the SM, one must take into account the corrections to the $\mu^2$ parameter, namely $\delta \mu^2$, induced by such new physics \cite{Farina:2013mla,deGouvea:2014xba}. The scale of new physics and how it couples to the SM Higgs sector determines the magnitude of $\delta \mu^2$ and the requirement that $\delta \mu^2$ should be of the order of $\mu^2$ (or TeV$^2$ in a more conservative approach \cite{deGouvea:2014xba}) implies constraints on the scale and couplings of new physics. In the case of type-I seesaw mechanism, one finds similar issue because of the existence of right handed neutrinos and their couplings to the SM fields\footnote{In general, the naturalness criteria gets modified in the presence of other new physics beyond the SM. In this paper, we however restrict ourselves to the study of type-I seesaw mechanism only.}. 
\begin{figure}[!ht]
\centering
\subfigure{\includegraphics[width=0.45\textwidth]{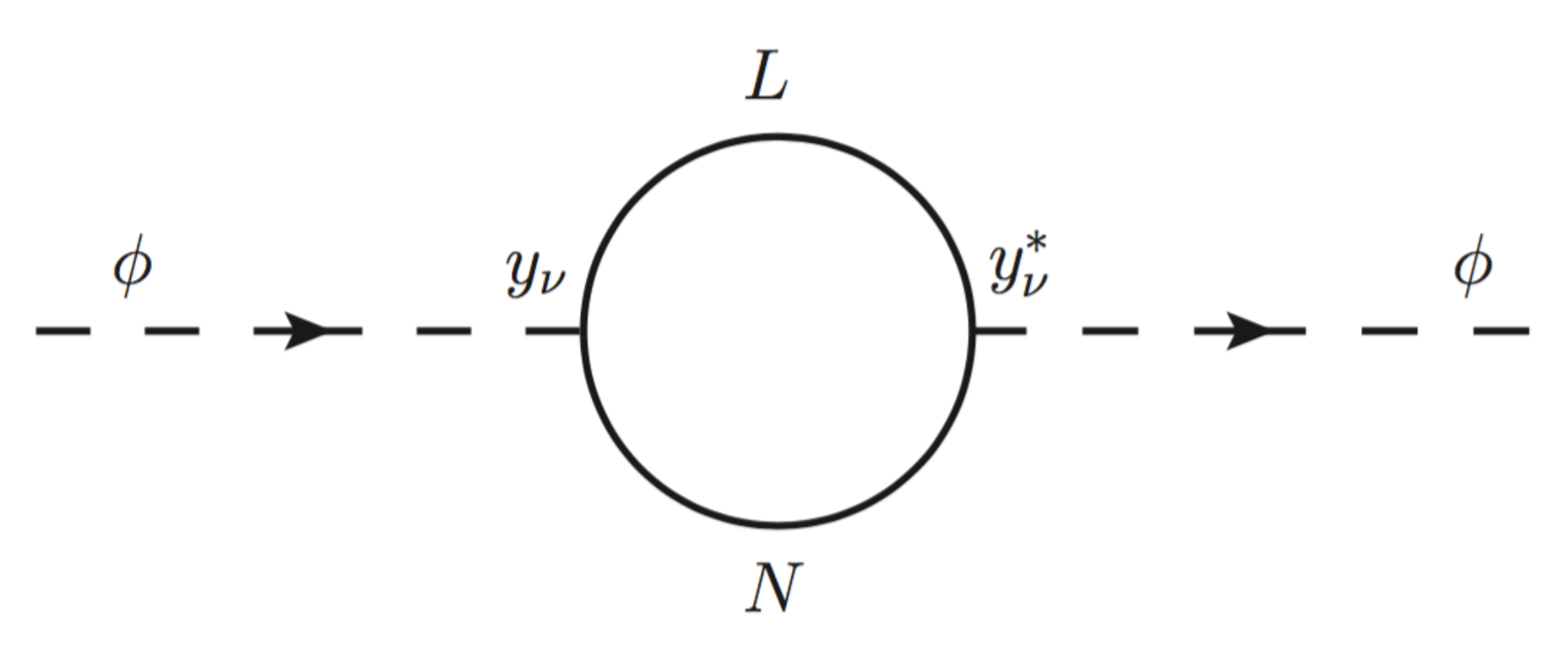}}
\caption{The one loop correction to $\mu^2$ parameter in the presence of Majorana fermion singlet $N$.}
\label{fig1}
\end{figure}
The one-loop correction to the renormalized $\mu^2$ parameter, represented by the Feynman diagram shown in Fig. \ref{fig1}, was first calculated by Vissani in \cite{Vissani:1997ys} in case of simple one-flavor type-I seesaw (see also \cite{Casas:2004gh,Xing:2009in,Davoudiasl:2014pya}). Such correction is estimated as 
\be \label{vissani}
\delta \mu^2 \approx \frac{1}{4 \pi^2}| y_\nu |^2 M_N^2 = \frac{1}{2 \pi^2} \frac{m_\nu}{v^2} M_N^3~, \ee
where $y_\nu$ is a coupling between the fermion singlet with SM leptons and Higgs and $M_N$ is mass of the fermion singlet. The $m_\nu = |y_\nu|^2 v^2/(2 M_N)$ is the seesaw mass for the SM neutrino. Clearly, $m_\nu = \sqrt{m_{\rm atm}} = 0.05$ eV leads to $M_N \le 10^7$ GeV if $\delta \mu^2$ is required to be smaller than (1 TeV)$^2$. Similar analysis for three flavoured type-I seesaw mechanism was carried out in \cite{Clarke:2015gwa} and it was found that at least two of the three fermion singlets are required to be lighter than $10^7$ GeV to maintain the electroweak naturalness. If the fermion singlets are close to the upper bound set by electroweak naturalness then it generically requires the Yukawa couplings of ${\cal O}(10^{-4})$ in order to produce viable light neutrino masses.  Even smaller Yukawa couplings are required when the masses of fermion singlets are further reduced.

In this paper we argue that type-I seesaw mechanism loses its inherent naturalness when the criteria of  elecroweak naturalness is imposed on it. The naturalness of both the Yukawa couplings and electroweak scale requires the masses of fermion singlets to be as light as of ${\cal O}$(TeV). The standard seesaw mechanism then can no longer be considered as the source of small neutrino masses. In this paper we modify the seesaw mechanism by incorporating it into the framework of discrete symmetries which give rise to massless SM neutrinos despite ${\cal O}(1)$ Yukawa couplings and TeV scale masses for the fermion singlets. Discrete symmetries have been widely used to predict flavour mixing patterns in the lepton sector, see \cite{Altarelli:2010gt,Altarelli:2012ss,Smirnov:2011jv,King:2013eh,
Ishimori:2010au} for some recent reviews. It is shown recently in \cite{Joshipura:2013pga,Joshipura:2014pqa,Joshipura:2014qaa,Joshipura:2015zla,Joshipura:2016hvn} that a class of discrete symmetries can also provide restrictions on the neutrino masses. We use this basic idea in order to suppress neutrino masses in electroweak natural seesaw setup. For this, we assume that the three generations of SM lepton doublets posses a global residual $Z_m \times Z_n \times Z_p$ symmetry with $m,n,p \ge 3$. This symmetry of SM neutrinos together with their Majorana nature implies all of them to be massless. The fermion singlets are assigned appropriate discrete symmetries in such a way that there exist three massive states of them and atleast one linear combination of fermion singlets couples with the SM leptons through Yukawa interactions. As we show in this paper, these conditions necessarily lead to two degenerate fermion singlets and one massive fermion singlet which completely decouples from the SM. The residual symmetries of leptons and fermion singlets can be combined into a discrete group $G_f$ which should be the symmetry of leptonic Lagrangian. The $G_f$ can be discrete subgroup (DSG) of SU(3) or U(3) depending on the representation to be chosen for the leptons. We provide specific model for each of this class of symmetries and discuss the phenomenology of generic perturbations which produces tiny neutrino masses. Our results open a new category of models in which viable neutrino masses in the low-scale seesaw frameworks are naturally realized through a mildly broken discrete symmetry.

The paper is organized as the following. In the next section, we revisit the constraints on type-I seesaw mechanism arising from the criteria of electroweak naturalness. In section \ref{sec:symmetry}, we formulate the general discrete symmetries which lead to the massless neutrinos through type-I seesaw despite ${\cal O}(1)$ Yukawa couplings and TeV scale fermion singlets. We consider specific examples of such symmetries in section \ref{sec:models}. The perturbations required to generate tiny neutrino masses are studied in section \ref{sec:perturbations}. Finally, we summarize in section \ref{sec:summary}.

\section{Electroweak naturalness and  type-I seesaw mechanism} 
\label{sec:ewnaturalness}
Consider an extension of the SM by $n$ number of gauge singlet Majorana fermions $N_\alpha$. Their complete renormalizable interactions can be written as
\be \label{L}
{\cal L} = {\cal L}_{SM}  + \overline{N}_\alpha \overline{\sigma}^\mu \partial_\mu N_\alpha - \frac{1}{2} {(M_N)}_{\alpha \beta} \overline{N^c}_\alpha N_\beta -  (Y_D)_{i\alpha} \overline{L}_i N_\alpha \tilde{\phi} + {\rm h.c.}~,\ee
where $i = 1,2,3$ and $\alpha,\beta=1,...,n$ are flavour indices, $L = (\nu_L, e_L)^T$, $\tilde{\phi}=i \tau_2 \phi^*$ and $\phi$ is the SM Higgs doublet with vacuum expectation value (VEV) of its electrically neutral component, $ \langle \phi \rangle \equiv v/\sqrt{2} = 174$ GeV. The $(Y_D)_{i \alpha}$ are the Dirac Yukawa couplings and $M_N$ is the Majorana mass matrix for heavy singlet fermions. Without loss of generality, one can consider a basis in which the $3\times 3$ charged lepton Yukawa matrix $Y_l$ and $M_N$ are diagonal with real and positive elements. We denote such diagonal elements in $M_N$ as $M_{N_\alpha}$. If $M_{N_\alpha} \gg \vev{\phi}$ then after the electroweak symmetry breaking, the SM neutrinos get the masses which can be expressed in terms of the fundamental couplings in eq. (\ref{L}) as
\be \label{mnu}
M_\nu = \vev{\phi}^2 Y_D M_N^{-1} Y_D^T~. \ee
The symmetric matrix $M_\nu$ is diagonalized by a unitary matrix such that
\be
U^\dagger M_\nu U^* = {\rm Diag.}(m_{\nu 1}, m_{\nu 2}, m_{\nu 3})~, \ee
where $m_{\nu i}$ are the masses of light neutrinos and $U$ is the leptonic mixing matrix, also known as the PMNS matrix. 

The Yukawa interaction of $N_\alpha$ with the light neutrinos and Higgs induces finite correction to the $\mu^2$ parameter at one-loop \cite{Vissani:1997ys,Casas:2004gh,Xing:2009in,Davoudiasl:2014pya}. Such a correction is estimated as
\be \label{dmu}
 |\delta \mu^2| \approx \frac{1}{4 \pi^2} \sum_{i,\alpha} |(Y_D)_{i\alpha}|^2 M_{N_\alpha}^2~.\ee
 The electroweak naturalness criteria therefore imposes a constraint
 \be \label{constraint}
 |(Y_D)_{i\alpha}| M_{N_\alpha} \le {\cal O} ({\rm TeV})~.\ee
It requires the singlet fermions at TeV scale if the neutrino Yukawa couplings are of ${\cal O}(1)$ or  small couplings if the mass scale of singlet fermion is heavier than TeV. The latter possibility is actually further constrained by the observed neutrino masses and one cannot consider arbitrarily small $(Y_D)_{i\alpha}$ and large $M_{N_\alpha}$. In a simplified case of single generation of light neutrino and fermion singlet, eq. (\ref{mnu}) implies $Y_D^2 \approx M_N m_\nu / \vev{\phi}^2$ leading to a generic bound from the criteria of naturalness:
\be \label{constraint_MN}
\frac{M_N^3 m_\nu}{4 \pi^2 \vev{\phi}^2} \lesssim {\rm (TeV)}^2 ~~~\Rightarrow ~~~ M_N \lesssim  2.9 \times 10^{7} \times \left( \frac{\sqrt{m_{\rm atm}^2}}{m_\nu}\right)^{1/3} ~{\rm GeV}~. \ee
The above bound on the mass of singlet fermion does not get drastically modified if three generations of neutrinos and fermion singlets are considered. A numerical investigation performed in \cite{Clarke:2015gwa} shows that all the three fermion singlets are generically required to be $\le 10^8$ GeV in order to produce viable neutrino masses and to maintain the electroweak naturalness. In a special case when the lightest neutrino is massless, one of the three singlets can have arbitrarily large mass unconstrained by the electroweak naturalness. This can also be understood from the fact that in such a case, a linear combination of the states $N_{\alpha}$ decouples completely from the SM and it has only the self interactions giving no contribution in the Higgs mass correction.

The electroweak naturalness demands the scale of fermion singlets below $10^{8}$ GeV. A typical observation from eq. (\ref{mnu}) then implies that the Dirac type Yukawa couplings are required to be small to account for light neutrino masses. If $m_i \le 0.1$ eV then $|(Y_D)_{i\alpha}| \lesssim {\cal O}(10^{-4})$. Further, if $M_{N_\alpha} \approx 1$ TeV then $|y_{i\alpha}| $ are typically required to be $ \lesssim 10^{-6}$. Such small couplings can arise from a more fundamental theory in which an underlying mechanism ensures the smallness of effective couplings. An example of such framework  is Froggatt-Neilsen based models in which an extra global U(1) symmetry and its spontaneous breaking is utilized to produce tiny effective couplings from the fundamental couplings of ${\cal O} (1)$ \cite{Altarelli:2002sg}. Another example is the theories based on extra spatial dimension in which the fermion singlets are localized far away from the SM brane (on which the Higgs is localized) leading to the small effective Dirac neutrino Yukawa couplings in four spacetime dimensions \cite{Kitano:2003cn,Feruglio:2014jla}. An alternative way to generate small neutrino masses with light fermion singlets and Yukawas of order unity is to have specific structures in $Y_D$ and $M_N$ such that $M_\nu$ vanishes in eq. (\ref{mnu}). This is termed as ``seesaw cancellation" and its phenomenology is studied in \cite{Kersten:2007vk,Dev:2013oxa}. The structures of $Y_D$ and $M_N$ leading to the seesaw cancellations remain very fine-tuned and unstable  with respect to higher order corrections if they are not consequences of some symmetry or a dynamical mechanism. One such framework is proposed in \cite{Kersten:2007vk} where the seesaw cancellation is shown to arise due to a global U(1) symmetry equivalent to the lepton number conservation. We offer an alternative framework in which seesaw cancellation arises from the residual discrete symmetries of the SM leptons and fermion singlets.

\section{Discrete symmetries and seesaw naturalness}
\label{sec:symmetry}
We provide a symmetry based origin of natural type-I seesaw in this section. When all the three SM neutrinos are strictly massless, the low energy effective theory obtained after integrating out the singlet fermions from eq. (\ref{L}) possesses a maximal accidental U(1)$^3$ $\equiv$ U(1)$_e$ $\times$ U(1)$_\mu$ $\times$ U(1)$_\tau$ global symmetry. The lepton doublets $L_{e,\mu,\tau}$ transform non-trivially under U(1)$_{e,\mu,\tau}$ and hence such symmetry leads to the same consequences as the lepton number conservation in the SM. Such a symmetry therefore can be used as guiding principle to forbid neutrino masses at the leading order in seesaw mechanism. The small perturbation are then induced in order to generate tiny neutrino masses. One such framework using a global U(1) symmetry and assuming three generations of fermion singlets is constructed in \cite{Kersten:2007vk}. The lepton doublets have $+1$ charge under this global symmetry while the charges of three singlet fermions are chosen appropriately such that all the Dirac Yukawa couplings do not vanish. It is shown that such a choice would always leads to one of the fermion singlets completely decoupled from the SM and the other two degenerate in masses which has non-vanishing Yukawa couplings with SM leptons and Higgs.

We adopt an alternate approach in which the suppression in neutrino masses originates from a discrete symmetry. The discrete symmetries are extensively used in order to predict the structure of leptonic mixing matrix. A different class of such symmetries can be utilized to predict particular mass patterns for Majorana neutrinos as it is recently shown in \cite{Joshipura:2013pga}. Appropriately chosen residual symmetry of neutrino mass matrix can lead to one massless neutrino \cite{Joshipura:2013pga,Joshipura:2014pqa} or two degenerate and one massive or massless neutrinos \cite{Joshipura:2014qaa}. A similar spectrum can be obtained by means of flavour antisymmetry \cite{Joshipura:2015zla,Joshipura:2016hvn}. Here, we extend this novel idea to all the three neutrinos and demand that such symmetry leads to  massless neutrinos by arranging appropriate cancellations in the seesaw formula eq. (\ref{mnu}). 

Consider a discrete flavour group $G_f$ as a symmetry of the leptonic part of Lagrangian in eq. (\ref{L}). Under $G_f$, the three generations of lepton doublets and fermion singlets  transform respectively as 
\be \label{transform}
L_i \to (S_L)_{ij} L_j~~{\rm and}~~N_i \to (S_N)_{ij} N_j~, \ee
where $i,j=1,2,3$ and sum over repeated index is implied. The $S_L$ and $S_N$ are $3\times 3 $ unitary matrices representing the transformation under the symmetry. An invariance of Lagrangian in eq. (\ref{L}) then implies 
\be \label{invariance}
S_L^\dagger Y_D S_N = Y_D~~{\rm and}~~S_N^T M_N S_N = M_N~. \ee
If there exists three massive fermion singlets then det.$S_N = \pm 1$. We choose det.$S_N = 1$ and hence $S_N$ as an element of DSG of SU(3) which is also a subgroup of underlying flavour group $G_f$. The most  general such $S_N$ in arbitrary basis is 
\be \label{SN}
S_N = V_N~{\rm Diag.}(\eta_1, \eta_2, \eta_1^*\eta_2^*)~V_N^\dagger~, \ee
where $V_N$ is unitary matrix representing arbitrary basis and $\eta_{1,2}$ are arbitrary phase factors. An invariance in eq. (\ref{invariance}) then implies 
\be \label{MN-form}
D_N \widetilde{M}_N D_N = \widetilde{M}_N~,\ee
where $\widetilde{M}_N = V_N^T M_N V_N$ and $D_N ={\rm Diag.}(\eta_1, \eta_2, \eta_1^*\eta_2^*) $. A requirement of three massive fermion singlets leads to two possibilities: (a) $\eta_{1,2}=\pm 1$ or (b) $\eta_1=\eta_2^* \neq \pm 1$. The choice (a) leads to three massive fermion singlets with no restrictions on their masses while (b) implies that two of the three fermion singlets are degenerate in masses. Such a symmetry is discussed earlier in \cite{Joshipura:2014qaa} in the context degenerate solar neutrino pair. 

It is easily seen from eq. (\ref{mnu}) and eq. (\ref{invariance}) that the effective neutrino mass matrix $M_\nu$ possesses residual symmetry such that 
\be \label{mnu-symm}
S_L^\dagger M_\nu S_L^* = M_\nu~. \ee
We assume that $S_L$ is an element of a group $Z_m \times Z_n \times Z_p$ with $m,n,p \ge 3$   such that it leads to all three massless neutrinos. In general, such an $S_L$ can be written as
\be \label{SL}
S_L = V_L~{\rm Diag.}(\zeta_1, \zeta_2, \zeta_3)~V_L^\dagger~, \ee
where $V_L$ is a unitary matrix and $\zeta_{1,2,3}$ are phase factors. The $M_\nu = 0$ requires $\zeta_{1,2,3}\neq \pm 1$ and $\zeta_i \zeta_j \neq 1$ for all $i\neq j$. Using this $S_L$ and $S_N$ from eq. (\ref{SN}), the symmetry constraints on the structure of Dirac Yukawa couplings can be determined from an invariance condition in eq. (\ref{invariance}). One obtains
\be \label{YD-sym}
D_L^* \widetilde{Y}_D D_N = \widetilde{Y}_D~, \ee
where $\widetilde{Y}_D = V_L^\dagger Y_D V_N$, $D_L={\rm Diag.}(\zeta_1, \zeta_2, \zeta_3) $ and $D_N$ as specified earlier. The matrix $\widetilde{Y}_D$ completely vanishes if $\eta_{1,2} = \pm 1$. Therefore the non-vanishing Dirac Yukawa couplings necessarily requires $\eta_1=\eta_2^* \equiv \eta \neq \pm 1$. The symmetry allowed by all three massive fermion singlets and non-vanishing Dirac Yukawa couplings therefore corresponds to
\be \label{SN-final}
S_N = V_N D_N V_N^\dagger~~{\rm with}~~D_N={\rm Diag.}(\eta,\eta^*,1)~~{\rm and}~~\eta\neq \pm1~. \ee
The $\widetilde{M}_N$ invariant under the above symmetry is
\be \label{structureMN}
\widetilde{M}_N=\left( \ba{ccc} 0 & M & 0 \\ M & 0 & 0 \\0 & 0 & M_3 \ea\right)~.\ee
It leads to a degenerate pair of Majorana fermions forming a psedo-Dirac state with mass $M$. It is also straightforward to see that the third column of $\widetilde{Y}_D$ vanishes and therefore the fermion singlet with mass $M_3$ decouples completely from the SM.

The structure of the first two columns of $\widetilde{Y}_D$ depends on the choice of phase factors. One gets non-vanishing element in the first (second) column and $j^{\rm th}$ row for $\eta \zeta_j^* = 1$  ($\eta \zeta_j = 1$). As it is discussed earlier, $\zeta_i \zeta_j \neq 1$ for any $i \neq j$ is required for $M_\nu = 0$ which implies that either the first or second column of $\widetilde{Y}_D$ must entirely vanish. Therefore, if \emph{all} the three SM neutrinos are arranged to couple with one fermion singlet then the following choices are allowed for $S_L$.
\be \label{SL-final}
S_L = V_L D_L V_L^\dagger~~{\rm with}~~D_L={\rm Diag.}(\eta,\eta,\eta)~~{\rm or}~~{\rm Diag.}(\eta^*,\eta^*,\eta^*)~. \ee
They respectively lead to 
\be \label{structureYD}
\widetilde{Y}_D = \left( \ba{ccc} \tilde{y}_1 & 0 & 0 \\ \tilde{y}_2 & 0 & 0 \\\tilde{y}_3 & 0 & 0 \ea\right)~{\rm or}~\left( \ba{ccc} 0 & \tilde{y}_1 &  0 \\ 0 &\tilde{y}_2 &  0 \\0 & \tilde{y}_3 &  0 \ea\right)~.\ee
In the mass basis of fermion singlets, one gets
\be \label{mass-basis-structure}
Y_D = \left( \ba{ccc} y_1 &  \pm i y_1 & 0 \\ y_2 &  \pm i y_2 & 0 \\y_3 & \pm i y_3 & 0 \ea\right)~~{\rm and}~~M_N={\rm Diag.}(M,M,M_3)~,\ee
where $y_i = \tilde{y}_i/\sqrt{2}$. Note that one can also choose $D_L = (\eta, \zeta_2, \zeta_3)$ with $\zeta_2 \neq \zeta_3^*$ and $\zeta_{2,3} \neq \pm 1$ and can obtain $\tilde{y}_2=\tilde{y}_3=0$ in the above $\widetilde{Y}_D$. These two cases are physically inseparable as both lead to $M_\nu =0$.

At this point we would like to compare our results with those obtained in \cite{Kersten:2007vk}. The same results have been obtained by the authors of \cite{Kersten:2007vk} enforcing the lepton number conservation without using discrete symmetries. We emphasize that the residual symmetry we use for the leptons, characterized by a generator $S_L$ given in eq. (\ref{SL-final}), can be seen as a DSG of U(1)$_{\rm L}$ global symmetry which corresponds to the lepton number conservation. Therefore while the basic mechanism to obtain the massless neutrinos is same, our approach offers an alternative way to realize seesaw cancellations through class of discrete symmetries. It therefore opens up a platform for discrete symmetry based model building for the electroweak natural seesaw.

\section{Models of natural seesaw based on discrete symmetries}
\label{sec:models}
We now provide some specific examples of $G_f$ which lead to massless neutrinos despite ${\cal O}(1)$ Yukawa couplings and low seesaw scale. As it is discussed in the previous section, such a $G_f$ must contain both $S_N$ and $S_L$, given in eq. (\ref{SN-final}) and eq. (\ref{SL-final}), as a symmetry of three generations of fermion singlets and lepton doublets respectively. If $S_N$ and $S_L$ both are simultaneously chosen to be diagonal, then it is sufficient to work with $G_f$ that has one-dimensional irreducible representations. Such a $G_f$ can be abelian group, $Z_n$ (with $n \ge 3$), with diagonal elements of $S_L$ and $S_N$ as its representations. In the simplest case, $G_f = Z_3$ is sufficient to generate the structure of $Y_D$ and $M_N$ given in eqs. (\ref{structureMN},\ref{structureYD}) if each of the three generations of lepton doublets and two of the three generations of fermion singlets transform non-trivially as one dimensional irreducible representations of $Z_3$. For example,  if $L_{1,2,3} \to \omega L_{1,2,3}$, $N_{1} \to \omega N_1$ and $N_2 \to \omega^2 N_2$ where $\omega = e^{2 \pi i/3}$ then one obtains $Y_D$ and $M_N$ as shown in eqs. (\ref{structureMN},\ref{structureYD}). Hence $Z_3$ is the smallest group which can be the symmetry of leptons leading to  the electroweak natural seesaw.

We however discuss more interesting class of symmetries under which either the three generations of $N_i$ or both $N_i$ and $L_i$ transform as three dimensional irreducible representations of an underlying group $G_f$. Let us first find out a suitable $G_f$ in which the three generations of fermion singlets can be assigned to a three dimensional irreducible representation. Such a group must contain $S_N = V_N {\rm Diag.}(\eta, \eta^*, 1) V_N^\dagger$ as one of its elements with $\eta \neq \pm 1$ and therefore a subgroup $Z_n$ with $n\ge 3$.  The smallest such group is $A_4$ which possesses two 3-dimensional and three 1-dimensional irreducible representations. The lepton doublets can be assigned to suitable 1-dimensional representations. We outline a complete model based on $A_4$ in the next subsection. 

If both $N_i$ and $L_i$ are chosen as 3-dimensional irreducible representations under a discrete group $G_f$ then such a group must contain both $S_N$ and $S_L$ given in eqs. (\ref{SN-final}) and (\ref{SL-final}) respectively. Since det.$S_L \neq 1$, such a group must be DSG of U(3) which is not a subgroup of SU(3). The DSG of U(3) containing at least one faithful three dimensional irreducible representation and of order upto 512 are listed in \cite{Ludl:2010bj}. We look for the groups which contain desired $S_L$ and $S_N$ as their elements. The smallest such group is found to be of order 81 and known as $\Sigma(81)$ in literature \cite{Hagedorn:2008bc,Ishimori:2010au,BenTov:2012xp}. We also construct a model based on this group and discuss it in the second subsection below.

\subsection{An $A_4$ Model}
\label{A4}
The group $A_4$ is the smallest DSG of SU(3) possessing 3-dimensional irreducible representation. This group has been widely used as a flavour symmetry for the leptons because of its ability to predict tri-bimaximal flavour mixing pattern in the lepton sector \cite{Altarelli:2010gt,Altarelli:2012ss,Smirnov:2011jv,King:2013eh,
Ishimori:2010au}. It has one 3-dimensional (${\bf 3}$) and three 1-dimensional (${\bf 1}$, ${\bf 1^\prime}$ and ${\bf 1^{\prime \prime}}$) irreducible representations. The tensor products and their decomposition rules are given in \cite{Ishimori:2010au}. We assume that the three flavours of fermion singlets transform as ${\bf 3}$ and each of the three flavours of lepton doublets transforms as ${\bf 1^\prime}$. The non-zero Dirac Yukawa couplings then require an existence of SM singlet scalar field $\chi = (\chi_1,\chi_2,\chi_3)^T$ which transform as ${\bf 3}$ under $A_4$. The gauge and $A_4$ invariant Lagrangian involving the leading order interactions of fermion singlets can be given as:
\be \label{A4-L}
-{\cal L}_N =  \frac{1}{\Lambda}y_i \overline{L}_i (N \chi)_{\bf 1^\prime} \tilde{\phi} + \frac{1}{2} M (\overline{N^c} N)_{\bf 1} +\frac{1}{2} \lambda (\overline{N^c} N)_{\bf 3} \chi + {\rm h.c.}~,\ee
where $(...)_{\bf r}$ denotes the component of the tensor product of the fields inside the bracket that transform as ${\bf r}$-dimensional irreducible representation.

Let's now discuss the breaking of $A_4$ symmetry induced by non-trivial VEV of flavon field $\chi$. In order to ensure that the mass matrix of fermion singlets remains invariant under $Z_3$ symmetry characterized by a generator similar to the one given in eq. (\ref{SN}), one needs to find such a generator in the representation of flavon field $\chi$ and demand that the vacuum of $\chi$ is invariant under the transformation induced by this generator. The generator of $Z_3$ subgroup of $A_4$ in the triplet representation is given by \cite{Ishimori:2010au}
\be \label{structureA4}
S_N = \left( \ba{ccc} 0 & 1 & 0 \\ 0 & 0 & 1 \\ 1 & 0 & 0 \ea\right)~. \ee
The constraint $S_N \vev\chi = \vev \chi$ implies the VEV structure $\vev{\chi_1}=\vev{\chi_2}=\vev{\chi_3}\equiv v_\chi$. It is discussed in detail in the Appendix B that such a VEV structure is naturally favoured for some range of parameters in the scalar potential.  After the $A_4$ symmetry is broken by the VEV, eq. (\ref{A4-L}) leads to 
\be \label{A4-result}
Y_D = \frac{v_\chi}{\Lambda}\left( \ba{ccc} y_1 & y_1 \omega & y_1 \omega^2 \\ y_2 & y_2 \omega & y_2 \omega^2 \\y_3 & y_3 \omega & y_3 \omega^2 \ea\right)
~~{\rm and}~~
M_N=\left( \ba{ccc} M & \lambda v_\chi & \lambda v_\chi \\ \lambda v_\chi & M & \lambda v_\chi \\ \lambda v_\chi & \lambda v_\chi & M \ea\right)~.\ee
The $Y_D$ and $M_N$ obtained in the above respect the constraints given in eq. (\ref{invariance}) and lead to massless neutrinos at the leading order. They can be brought into the form given in eq. (\ref{mass-basis-structure}) through a basis transformation 
\be \label{}
M_N \to U^T M_N U~, ~~Y_D \to Y_D U,~~{\rm with}~~U=\left( \ba{ccc} \sqrt{\frac{2}{3}} & 0 & \frac{1}{\sqrt{3}} \\
-\frac{1}{\sqrt{6}} & -\frac{1}{\sqrt{2}} & \frac{1}{\sqrt{3}} \\
-\frac{1}{\sqrt{6}} & -\frac{1}{\sqrt{2}} & \frac{1}{\sqrt{3}}\ea\right)~.\ee

\subsection{A $\Sigma(81)$ Model}
\label{81}
We now discuss the model in which both the fermion singlets and lepton doublets can be assigned 3-dimensional irreducible representations. The group $\Sigma(81)$ has eight triplets (${\bf 3}_A$, ${\bf 3}_B$, ${\bf 3}_C$, ${\bf 3}_D$,  ${\bf \bar{3}}_A$, ${\bf \bar{3}}_B$, ${\bf \bar{3}}_C$ and ${\bf \bar{3}}_D$) and nine singlets (${\bf 1}^k_l$ with $k,l=0,1,2$) \cite{Ishimori:2010au}. The generators represented on each of the triplets are listed in \cite{Ishimori:2010au} which we reproduce in the Appendix A for a convenience of reader. A complete set of tensor product decomposition rules are also listed in the Appendix A. We assign ${\bf 3}_D$ (${\bf 3}_C$) representation to the three flavours of fermion singlets (lepton doublets). We require four flavon fields to reproduce completely the ansatz given in eq. (\ref{mass-basis-structure}). They are denoted as $\varphi \sim {\bf 3}_D$, $\psi_A \sim {\bf 3}_A$, $\psi_B \sim {\bf 3}_B$ and $\psi_C \sim {\bf 3}_C$. The relevant part of the Lagrangian of  model at the leading order is 
\beqa \label{S81-L}
-{\cal L}_N  &=&  \frac{1}{\Lambda}y (\overline{L} N)_{\overline{\bf 3}_A} \psi_A \tilde{\phi} + \frac{1}{\Lambda}y^\prime (\overline{L} N)_{\overline{\bf 3}_B} \psi_B \tilde{\phi} + \frac{1}{\Lambda}y^{\prime \prime} (\overline{L} N)_{\overline{\bf 3}_C} \psi_C \tilde{\phi}  \nonumber \\
&+&  \frac{1}{2} \lambda (\overline{N^c} N)_{\overline{\bf 3}_D} \varphi +\frac{1}{2} \lambda^\prime (\overline{N^c} N)_{\overline{\bf 3}_D} \varphi  + {\rm h.c.}~,\eeqa
where $(\overline{N^c} N)_{\overline{\bf 3}_D}$ in the last two terms represent two different invariant combinations of the product $\overline{N^c} N$ as listed in the tensor  decomposition rule eq. (\ref{3d3d}) given in Appendix A. 

The $\Sigma(81)$ symmetry is to be broken down to the $Z_3$ symmetry corresponding to the generator $a^\prime$, represented on ${\bf 3}_D$ as given in eq. (\ref{3dgen}), in the fermion singlet sector. The VEV of $\varphi$  therefore must respect $a^\prime \vev \varphi = \vev \varphi$. This implies 
\be\label{81-varphi-vac}
\vev{\varphi_1} =\vev{\varphi_2} =0,~\vev{\varphi_3} \equiv v_\varphi \neq 0. \ee
We further require the Dirac Yukawa couplings to be invariant under the symmetry transformation given in eq. (\ref{invariance}) with $S_L = a a^\prime a^{\prime \prime} = {\rm Diag.}(\omega, \omega, \omega)$. Using the tensor product decomposition rules we have derived for ${\bf \bar{3}}_C \otimes {\bf 3}_D$ given in eq. (\ref{3cb3d}), we find that the VEVs of flavons must follow $(a {a^\prime}^2)_{{\bf 3}_A} \vev{\psi_A} = \vev{\psi_A}$, $(a^2 a^{\prime \prime})_{{\bf 3}_B} \vev{\psi_B} = \vev{\psi_B}$ and $(a^\prime {a^{\prime \prime}}^2)_{{\bf 3}_C} \vev{\psi_C} = \vev{\psi_C}$, where the $(...)_{\bf r}$ implies the generators in the parenthesis to be chosen in ${\bf r}^{\rm th}$ representation. These constraints lead to 
\be \label{vacuum-81}
\vev{\psi_A} = v_{\psi_A} \left( \ba{c} 0 \\ 0 \\ 1 \ea \right)~,~~~\vev{\psi_B} = v_{\psi_B} \left( \ba{c} 0 \\ 1 \\ 0 \ea \right)~,~~\vev{\psi_C} = v_{\psi_C} \left( \ba{c} 1 \\ 0 \\ 0 \ea \right)~.
\ee
The resulting structures of $Y_D$ and $M_N$ when compared to those in eq. (\ref{L}) are
\be \label{S81-result}
Y_D = \frac{1}{\Lambda}\left( \ba{ccc} y v_{\psi_A}   & 0 & 0 \\ y^{\prime \prime} v_{\psi_C} & 0 & 0 \\y^\prime v_{\psi_B} & 0 & 0\ea\right)
~~{\rm and}~~
M_N= v_\varphi \left( \ba{ccc} 0 & \lambda & 0\\ \lambda & 0 & 0 \\ 0 & 0 & \lambda^\prime \ea\right)~.\ee
Again, they can be brought into the form given in eq. (\ref{mass-basis-structure}) by the basis transformation 
\be \label{}
M_N \to U^T M_N U~, ~~Y_D \to Y_D U,~~{\rm with}~~U=\left( \ba{ccc} \frac{1}{\sqrt{2}} & -\frac{i}{\sqrt{2}} &0\\
\frac{1}{\sqrt{2}} & \frac{i}{\sqrt{2}} &0 \\
0 & 0 & 1\ea\right)~,\ee
resulting into $y_1=\frac{y}{\sqrt{2}} \frac{v_{\psi_A}}{\Lambda}$, $y_2= \frac{y^{\prime \prime}}{\sqrt{2}} \frac{v_{\psi_C}}{\Lambda}$, $y_3=\frac{y^{\prime}}{\sqrt{2}} \frac{v_{\psi_B}}{\Lambda}$, $M=\lambda v_\varphi$ and $M_3=\lambda^\prime v_\varphi$. Clearly, this model requires more flavons and therefore is less economical than the model based on $A_4$ symmetry discussed earlier. We also discuss the viability of VEV structures in Appendix B. It is shown that the above vacuum structure is not natural in the given minimal model. To obtain the required vacuum alignment without any fine tunning in the scalar potential, one needs to suitably modify the model. We have disccussed one such possibility in Appendix B in which an additional $U(1)$ symmetry is imposed under which all the flavon fields posses different charges. However a set of new flavons, charged under the $U(1)$ but singlet under $\Sigma(81)$, is required to maintain the form of interactions in Eq. (\ref{S81-L}). We refer reader to Appendix B for more details.

One of the motivations to assign 3-dimensional irreducible representations to the leptons is to predict their flavour mixing pattern through underlying symmetry. However, in the present case all the mixing angles are not physical as the neutrino masses are vanishing at the leading order. So the presence of discrete symmetry here does not correspond to any prediction for the mixing angles. Once the symmetry is broken to generate tiny neutrino masses, it gives rise to the leptonic mixing parameters. We quantitatively discuss some generic cases of the symmetry breaking in the next section.

Before ending this section, we comment on the viability of the above models in the context of electroweak naturalness. Both the proposed models involve non-renormalizable interactions as well as new SM singlet flavon fields. For example, such a flavon field $\xi$ can couple to the SM Higgs $\phi$ with a coupling like
\be \label{flavon-higgs}
{\cal L}_{\phi-\xi } = \kappa |\phi|^2 |\xi|^2~, \ee
which is not forbidden by the SM gauge or flavour symmetry. This interaction contributes in the $\mu^2$ correction at one-loop level which is estimated to be \cite{deGouvea:2014xba}
\be \label{}
\delta \mu^2 \sim \frac{\kappa}{16 \pi^2} M_\xi^2~,\ee
where $M_\xi$ represents the mass of flavon field. The elecroweak naturalness then requires either $\kappa \ll 1$ or $M_\xi \le 1$ TeV. This implies that such flavons should be present at TeV scale in the natural theories. Moreover the models presented in the above involve non-renormalizable interactions. The ultra-violate completions of these effective interactions often require presence of additional vector-like leptons and/or new scalar fields. Since these new leptons do have non-zero SM gauge quantum numbers, they also contribute in the Higgs mass corrections. If all the dimensionless couplings are taken to be ${\cal O}(1)$ then electroweak naturalness again dictates the mass scale of such new fields to be ${\cal O}$(TeV) \cite{deGouvea:2014xba}. Hence one finds the scale of discrete symmetry breaking and the cutoff scale $\Lambda$ to be close and $\sim {\cal O}$(TeV) in natural theories. If the criteria of naturalness is given up and low scale type-I seesaw mechanism is still considered then the scale of discrete symmetry breaking and  $\Lambda$ can be arbitrarily large keeping the ratio $\vev{\xi}/\Lambda$ to be of ${\cal O}(1)$.

\section{Breaking of residual symmetries \& nonzero neutrino masses}
\label{sec:perturbations}
We now discuss generic perturbations in $Y_D$ and $M_N$ given in eq. (\ref{mass-basis-structure}) derived by demanding invariance under the residual symmetries $S_L$ and $S_N$. Such perturbations may arise from different sources depending on the exact model under consideration. The most common source of perturbation is the next-to-leading order corrections in $Y_D$ and/or $M_N$ which do not respect the invariance conditions eq. (\ref{invariance}). Another source of perturbation is a small deviation from the exact vacuum alignments in the flavon fields which may arise again from the next-to-leading order corrections in the flavon potential. We however do not discuss here the origin of such model specific perturbations and only analyze phenomenological consequences of generic perturbations. In the mass basis of fermion singlets, the most general deviations form $Y_D$ and $M_N$ in eq. (\ref{mass-basis-structure}) can be parametrized as 
\be \label{perturbations}
Y_D^\prime = \left( \ba{ccc} y_1  (1 - \epsilon_1) &   i y_1 (1 + \epsilon_1) & \epsilon_4 \\ y_2  (1 - \epsilon_2) &  i y_2 (1 + \epsilon_2) & \epsilon_5 \\y_3  (1 - \epsilon_3) &   i y_3 (1 + \epsilon_3) & \epsilon_6 \ea\right)~,~~M_N^\prime={\rm Diag.}(M  (1 - \epsilon_M),M  (1 + \epsilon_M),M_3)~.\ee

\begin{figure}[!ht]
\centering
\subfigure{\includegraphics[width=0.48\textwidth]{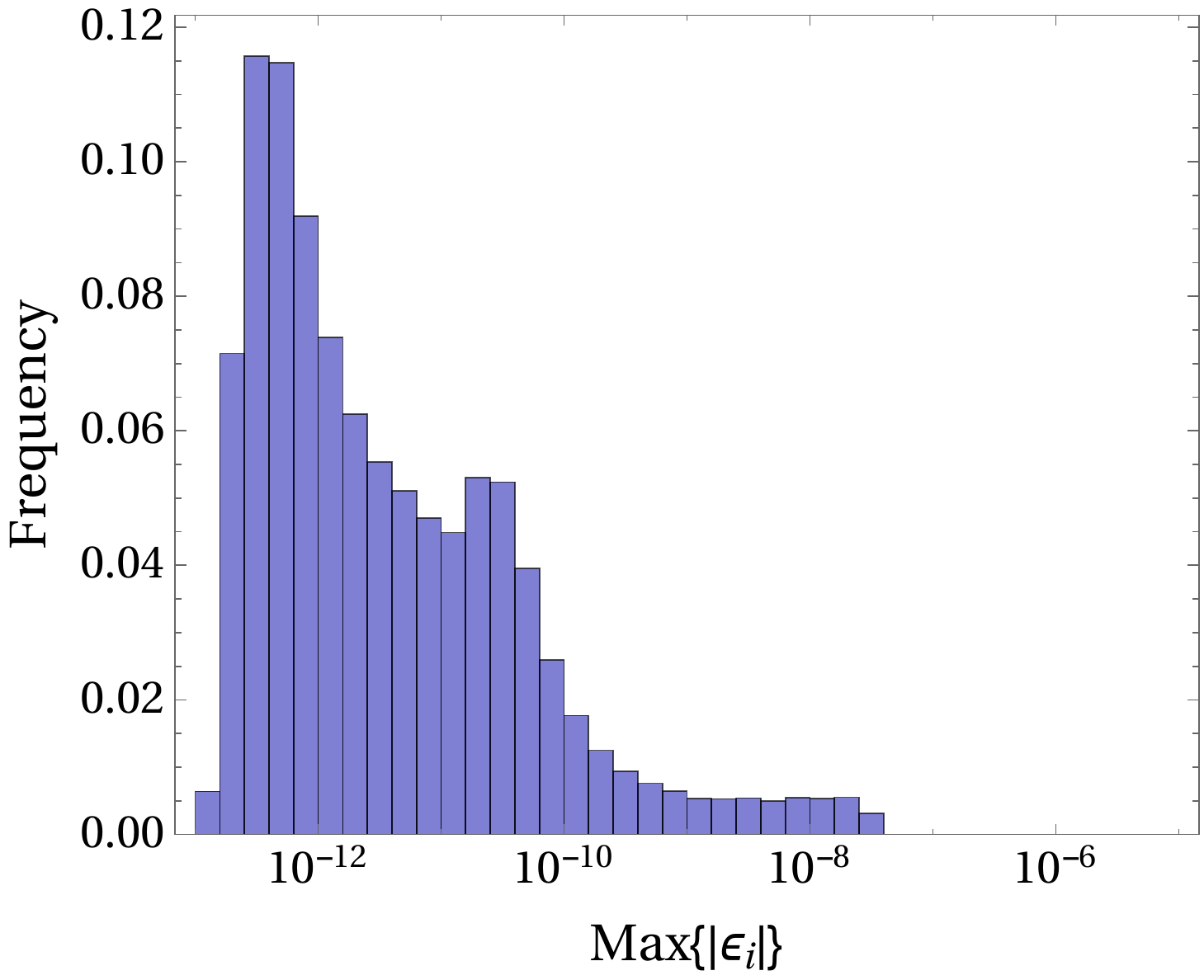}}
\caption{The largest magnitude of perturbation required in eq. (\ref{perturbations}), with $\epsilon_{4,5,6} = \epsilon_M=0$, in order to generate $\Delta m^2_{\rm sol}$ and $\Delta m^2_{\rm atm}$ within the $3\sigma$ of their global fit values \cite{Esteban:2016qun}. The values of $y_i$ are taken from the random and flat distribution corresponding to the range $|y_i| \in [0.5,1.5]$ and arg.$(y_i) \in [0,2 \pi]$ and $M_3 = 2M$ is fixed at 2 TeV.}
\label{fig2}
\end{figure}

We discuss two phenomenologically interesting cases. In the first case, we assume that the mass matrix of fermion singlets still possesses suitable residual $Z_n$ symmetry characterized by $S_N$ given in eq. (\ref{SN-final}) while the Dirac Yukawa interactions do not respect such symmetry {\it i.e.} $S_L^\dagger Y_D S_N \neq Y_D$. This case is characterized by $\epsilon_M = 0$ in eq. (\ref{perturbations}). Such scenario may arise, for example in case of $\Sigma(81)$ model when the vacuum of flavons $\psi_{A,B,C}$ has small deviations from their  structures given in eq. (\ref{vacuum-81}) but the VEV alignment of $\varphi$ remains intact. To analyze this case, we fix $M_3 = 2M = 2$ TeV and randomly vary all $y_i$ in the range: $|y_i| \in [0.5,1.5]$ and arg$(y_i) \in [0,2 \pi]$. For each of these point, we optimize the values of $\epsilon_i$ such that they reproduce the solar and atmospheric squared mass differences within the $3\sigma$ ranges of their global fit values. These ranges are taken from the recent global fit of the neutrino oscillation data given in \cite{Esteban:2016qun}. Here, we do not impose any restrictions on $\epsilon_i$ from the neutrino mixing angles since the mixing angles depend also on the parameters in the charged lepton mass matrix which, in general case, is also perturbed by symmetry breaking effects. We first set $\epsilon_{4,5,6} = 0$ in eq. (\ref{perturbations}) which corresponds to a case with perturbations but still with decoupled fermion singlet corresponding to the mass $M_3$. The results of this case are displayed in Fig. \ref{fig2}. 
It can be seen that one requires very small perturbations $\lsim {\cal O}(10^{-8}-10^{-13})$ in order to produce viable neutrino mass spectrum. The resulting neutrino mass spectrum has a massless neutrino as one of the fermion singlets completely decouples from the SM. 
\begin{figure}[!ht]
\centering
\subfigure{\includegraphics[width=0.48\textwidth]{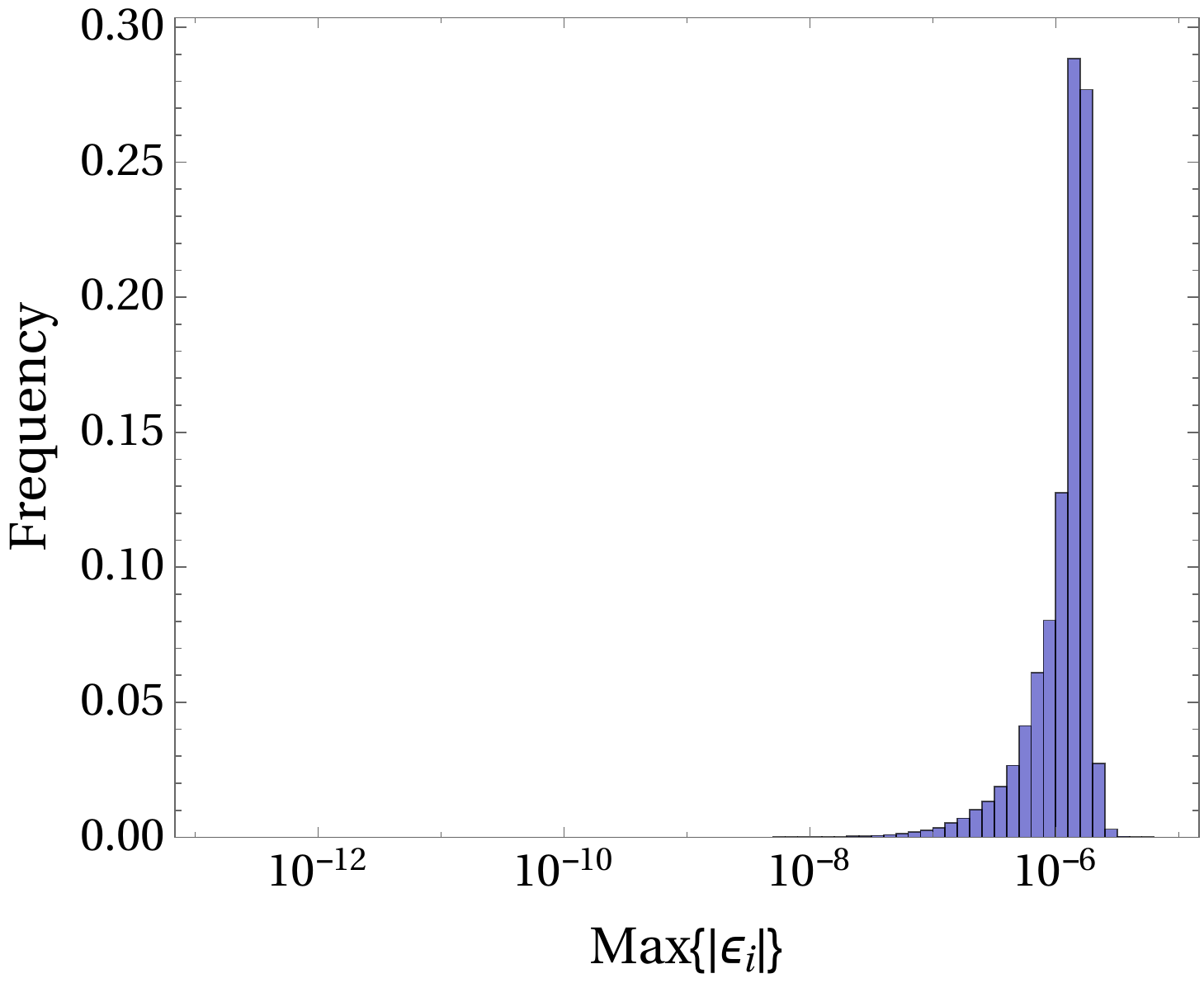}}\quad
\subfigure{\includegraphics[width=0.48\textwidth]{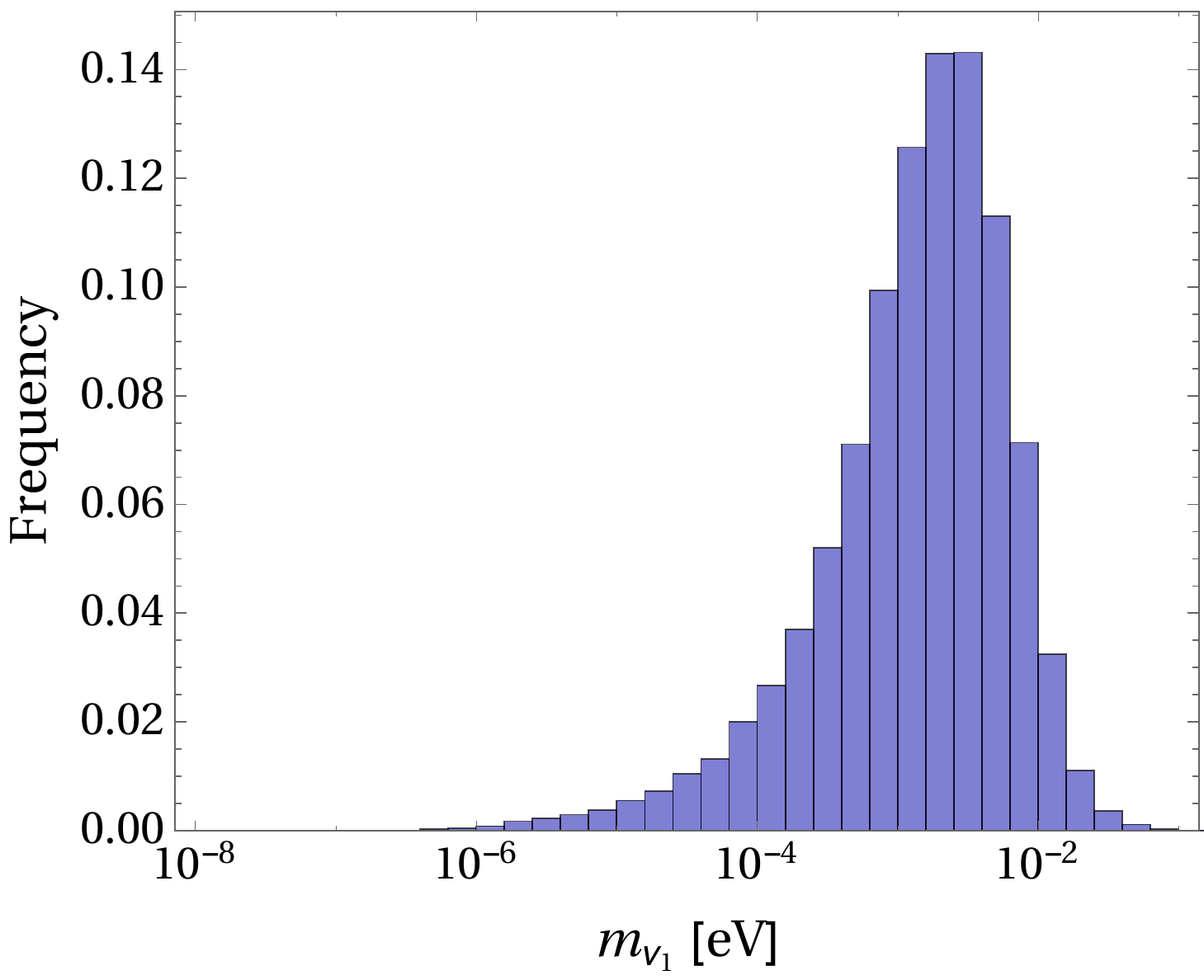}}\\
\caption{Left: Same as the caption of Fig. \ref{fig2} but with $\epsilon_{4,5,6} \neq 0$. Right: The corresponding predictions for the mass of the lightest neutrino..}
\label{fig3}
\end{figure}
In Fig. \ref{fig3}, we show the similar results but with $\epsilon_{4,5,6} \neq 0$. As it can be seen, in this case the magnitude of perturbation can be as large as ${\cal O}(10^{-6})$ and the lightest neutrino can be heavier compared to the previous case.

\begin{figure}[!ht]
\centering
\subfigure{\includegraphics[width=0.48\textwidth]{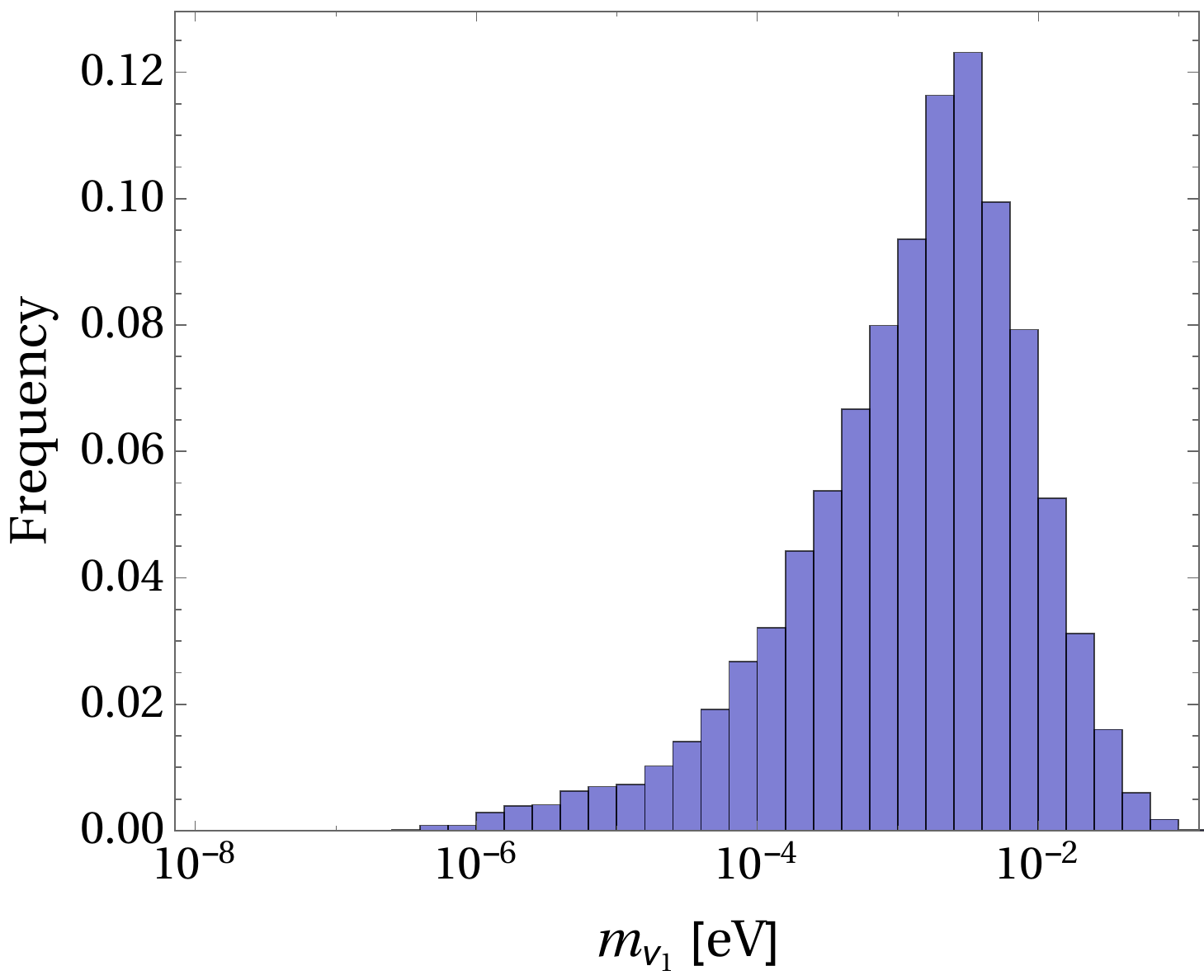}}\quad
\subfigure{\includegraphics[width=0.48\textwidth]{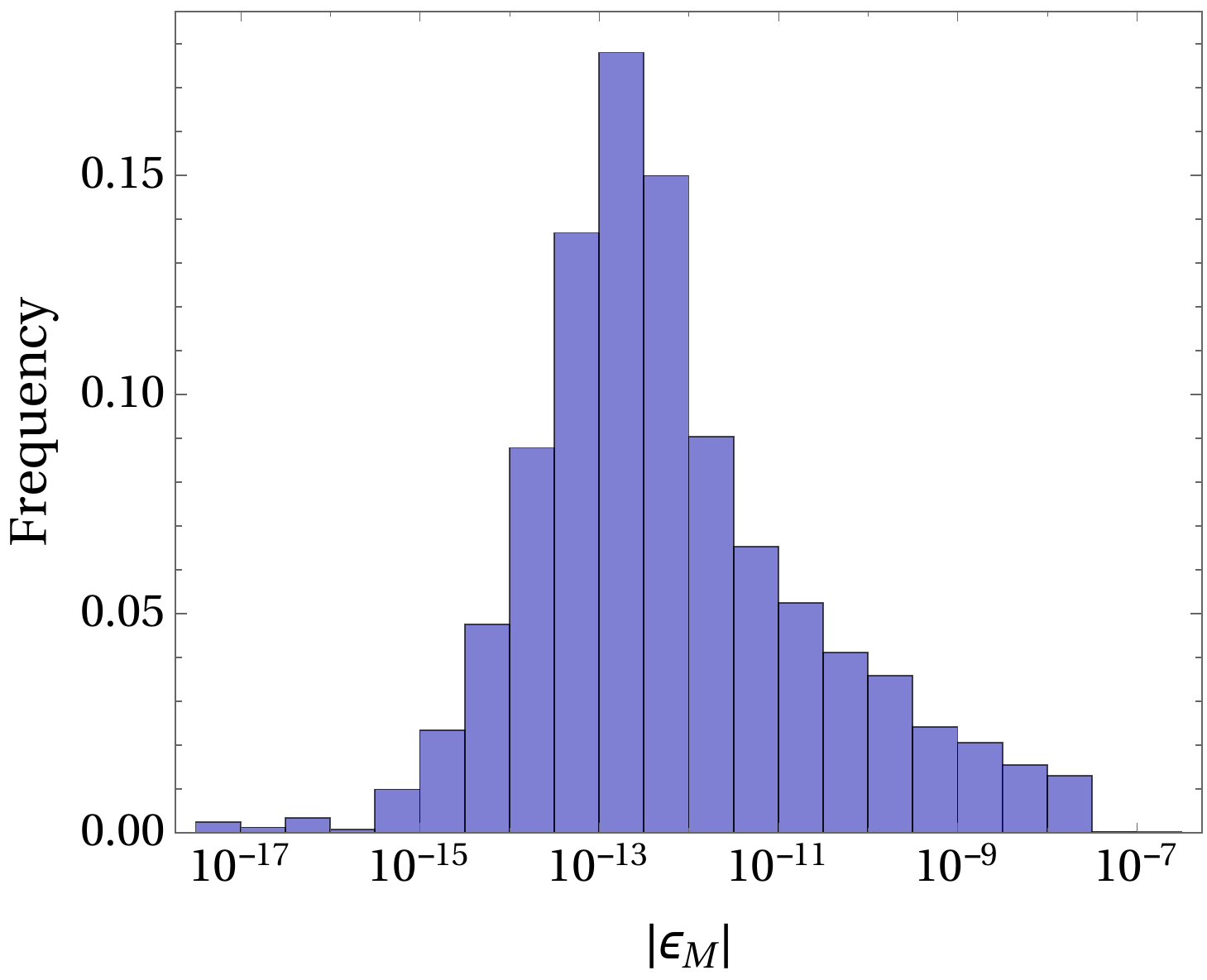}}\\
\caption{The predictions for the mass of the lightest neutrino (left) and splittings between the masses of fermion singlets of the first and second generations (right) allowed by the most general perturbation in eq. (\ref{perturbations}). The other details are same as mentioned in the caption of Fig. \ref{fig2}.}
\label{fig4}
\end{figure}

In the second case, we assume that perturbation breaks both the residual symmetries and both $Y_D$ and $M_N$ do not satisfy the invariance conditions given in eq. (\ref{invariance}). In this case, $\epsilon_M$ can also be nonzero together with all the $\epsilon_i$ in $Y_D$. We find that the magnitude of $\epsilon_{3,4,5}$ dominate over all the other perturbations. The largest magnitude of perturbations required in this case is similar to the one shown in the left panel of Fig. \ref{fig3}. The other results of this case are displayed in Fig. \ref{fig4}. We find that the allowed splitting between the fermion singlets of first two generations varies in the range $10^{-3}$-$10^5$ eV for $M=1$ TeV. Although we assume normal ordering for the neutrino masses in the above numerical analysis of perturbations, we get similar results in the case of inverted ordering. 

\section{Summary and Discussion}
\label{sec:summary}
In the models of neutrino masses based on type-I seesaw mechanism, the naturalness of electroweak scale restricts the masses of fermion singlets to be $\le 10^{7}$ GeV and their Yukawa couplings to the SM fermion to be of ${\cal O}(10^{-4})$.  If these couplings are assumed to be of the order of unity then fermion singlets are required to be as light as few TeV. In this case, the seesaw mechanism cannot be considered as the dominant mechanism responsible for small neutrino masses. In order to produce phenomenologically viable neutrino mass spectrum in this setup, one needs specific finely tuned correlations among the ${\cal O}(1)$ Yukawa couplings and masses of singlet fermions. We motivate such fine-tuning through the presence of finite discrete symmetry under which all the SM leptons and fermion singlets transform non-trivially. The three generations of lepton doublets are assumed to possess $Z_m\times Z_n \times Z_p$ symmetry with $m,n,p \ge 3$ which lead to massless neutrinos at the leading order. The transformation rules of three generations of singlet fermions under the action of symmetry are then suitably chosen such that it leads to three massive Majorana fermion singlets and atleast one linear combination of them coupling to the SM leptons. These requirements necessarily lead to a pair of degenerate fermion singlets irrespective of the underlying symmetry group. 

One can choose the discrete symmetry group $G_f$ depending on the representations to be assigned to the leptons and fermion singlets. The $G_f$ can be abelian symmetry if the one dimensional representations are chosen for the leptons and fermion singlets. If 3-dimensional irreducible representation is assigned to the three generations of fermion singlets, then one can have $G_f$ as a DSG of SU(3). The smallest such group is $A_4$ and we have provided a model realization of it. If the leptons are also to be chosen as 3-dimensional irreducible representation then the $G_f$ is necessarily DSG of U(3). We find the group $\Sigma(81)$ as the smallest DSG of U(3) which qualifies to be a symmetry of massless neutrinos and we have outlined a model based on this group. It is found that $A_4$ symmetry provides more economical and natural option for model compared to the $\Sigma(81)$. In all the cases, the underlying symmetry leads to massless neutrinos at the leading order and the tiny neutrino masses arise through small perturbations to the symmetry. We also study the phenomenology of generic perturbations and find the magnitude of such perturbations required to generate viable neutrino masses.  We find that small deviations from degeneracy in the masses of fermion singlets is compatible with the data and therefore the resonant leptogenesis mechanism may naturally emerge in this class of models as an alternative of the standard thermal leptogenesis. It is also possible to have the similar symmetry based realization for seesaw cancellations in case of only two generations of fermion singlets. One still gets the degenerate pair of fermion singlets at the leading order by the symmetry conditions. This setup naturally leads to the minimal low scale type-I seesaw model with (quasi)degenerate RH neutrinos and its updated phenomenology is recently studied in \cite{Bambhaniya:2016rbb, Rink:2016knw}. It is shown there that such a framework can successfully account for the baryon asymmetry of the universe through resonant leptogenesis.

\begin{acknowledgments} We thank Anjan S. Joshipura for reading the manuscript and useful suggestions. KMP thanks the Department of Science and Technology, Government of India for research grant support under INSPIRE Faculty Award (DST/INSPIRE/04/2015/000508). We acknowledge the use of JaxoDraw \cite{Binosi:2008ig}.
\end{acknowledgments}

\appendix
\section{The group $\Sigma(81)$}
Here we outline some important features of the group $\Sigma(81)$. The reader is advised to see \cite{Ishimori:2010au} for more details. The $\Sigma(81)$ group is a finite discrete subgroup of U(3) which has 81 elements. The elements can be written as $g=b^k a^ l a^{\prime m} a^{\prime \prime n}$ with $k,l,m,n=0,1,2$. The generators $b,a,a^\prime$ and $a^{\prime \prime}$ satisfy $b^3=a^3=a^{\prime 3}=a^{\prime \prime 3} = 1$, $b^{-1}ab = a^{\prime \prime}$, $b^{-1} a^\prime b = a$ and $b^{-1}a^{\prime \prime} b = a^\prime$. The generators $a$, $a^\prime$ and $a^{\prime \prime}$ commute with each others. The elements are classified into seventeen conjugacy classes. There are nine singlets represented by $\textbf{1}^k_l$ where $k,l=0,1,2$ and eight triplets represented by $\textbf{3}_A$, $\textbf{3}_B$, $\textbf{3}_C$, $\textbf{3}_D$, $\overline{\textbf{3}}_A$, $\overline{\textbf{3}}_B$, $\overline{\textbf{3}}_C$ and $\overline{\textbf{3}}_D$. Below we list the set of four generators for each of these 3-dimensional representations.

On all of the triplets, the generator b is represented as
\be
b=\begin{pmatrix}
0 & 1& 0\\
0 & 0 & 1\\
1& 0& 0
\end{pmatrix}~.
\ee\\

The representation of the generators $a$, $a^\prime$ and $a^{\prime \prime}$ on each of the triplets are:\\
\be
a=\begin{pmatrix}
\omega & 0& 0\\
0& 1& 0\\
0& 0& 1
\end{pmatrix},~
a^\prime =\begin{pmatrix}
1& 0& 0\\
0& \omega & 0\\
0& 0& 1
\end{pmatrix},~
a^{\prime \prime}=\begin{pmatrix}
1& 0& 0\\
0& 1& 0\\
0& 0& \omega
\end{pmatrix},~~~~{\text{on~$\textbf{3}_A$}}~\ee \\
\be
a=\begin{pmatrix}
1& 0& 0\\
0& \omega^2& 0\\
0& 0& \omega^2
\end{pmatrix},~
a^\prime=\begin{pmatrix}
\omega^2& 0& 0\\
0& 1& 0\\
0& 0& \omega^2
\end{pmatrix},~
a^{\prime \prime}=\begin{pmatrix}
\omega^2& 0& 0\\
0& \omega^2& 0\\
0& 0& 1
\end{pmatrix}, ~~~~{\text{on $\textbf{3}_B$}}~\ee \\
\be 
a=\begin{pmatrix}
\omega^2& 0& 0\\
0& \omega& 0\\
0& 0& \omega
\end{pmatrix},~
a^\prime=\begin{pmatrix}
\omega& 0& 0\\
0& \omega^2& 0\\
0& 0& \omega
\end{pmatrix},~
a^{\prime \prime}=\begin{pmatrix}
\omega& 0& 0\\
0& \omega& 0\\
0& 0& \omega^2
\end{pmatrix},~~~~{\text{on $\textbf{3}_C$}}~\ee \\
\be \label{3dgen}
a=\begin{pmatrix}
\omega^2& 0& 0\\
0& 1& 0\\
0& 0& \omega
\end{pmatrix},~
a^\prime=\begin{pmatrix}
\omega& 0& 0\\
0& \omega^2& 0\\
0& 0& 1
\end{pmatrix},~
a^{\prime \prime}=\begin{pmatrix}
1& 0& 0\\
0& \omega& 0\\
0& 0& \omega^2
\end{pmatrix},~~~~{\text{on $\textbf{3}_D$}}. \ee
The representations on $\overline{\textbf{3}}_A$, $\overline{\textbf{3}}_B$, $\overline{\textbf{3}}_C$, $\overline{\textbf{3}}_D$ are complex conjugate of the representations of $\textbf{3}_A$, $\textbf{3}_B$, $\textbf{3}_C$, $\textbf{3}_D$ respectively. The generators are represented on the singlets $\textbf{1}^k_l$ as $b=\omega^l$, $a=a^\prime=a^{\prime \prime}=\omega^k$.

We now list the complete set of tensor product decompositions for the triplets. 
\begin{equation}
\begin{pmatrix}
a_1\\
a_2\\
a_3
\end{pmatrix}_{\textbf{3}_A} \otimes
\begin{pmatrix}
b_1\\
b_2\\
b_3
\end{pmatrix}_{\textbf{3}_A}=
\begin{pmatrix}
a_1b_1\\
a_2b_2\\
a_3b_3
\end{pmatrix}_{\overline{\textbf{3}}_A}\oplus
\begin{pmatrix}
a_2b_3\\
a_3b_1\\
a_1b_2
\end{pmatrix}_{\overline{\textbf{3}}_B}\oplus
\begin{pmatrix}
a_3b_2\\
a_1b_3\\
a_2b_1
\end{pmatrix}_{\overline{\textbf{3}}_B}
\end{equation}\\

\begin{equation}
\begin{split}
\begin{pmatrix}
a_1\\
a_2\\
a_3
\end{pmatrix}_{\textbf{3}_A}\otimes
\begin{pmatrix}
b_1\\
b_2\\
b_3
\end{pmatrix}_{\overline{\textbf{3}}_A}=
\Big(\sum\limits_{l=0,1,2}(a_1b_1+\omega^{2l}a_2b_2+\omega^la_3b_3)_{\textbf{1}_l^0}\Big)\\\oplus
\begin{pmatrix}
a_2b_1\\
a_3b_2\\
a_1b_3
\end{pmatrix}_{\textbf{3}_D}\oplus
\begin{pmatrix}
a_1b_2\\
a_2b_3\\
a_3b_1
\end{pmatrix}_{\overline{\textbf{3}}_D}
\end{split}
\end{equation}\\

\begin{equation}
\begin{pmatrix}
a_1\\
a_2\\
a_3
\end{pmatrix}_{\textbf{3}_A}\otimes
\begin{pmatrix}
b_1\\
b_2\\
b_3
\end{pmatrix}_{\textbf{3}_B}=
\begin{pmatrix}
a_1b_1\\
a_2b_2\\
a_3b_3
\end{pmatrix}_{\overline{\textbf{3}}_C}\oplus
\begin{pmatrix}
a_3b_2\\
a_1b_3\\
a_2b_1
\end{pmatrix}_{\overline{\textbf{3}}_A}\oplus
\begin{pmatrix}
a_2b_3\\
a_3b_1\\
a_1b_2
\end{pmatrix}_{\overline{\textbf{3}}_A}
\end{equation}\\

\begin{equation}
\begin{split}
\begin{pmatrix}
a_1\\
a_2\\
a_3
\end{pmatrix}_{\textbf{3}_A}\otimes
\begin{pmatrix}
b_1\\
b_2\\
b_3
\end{pmatrix}_{\overline{\textbf{3}}_B}=
\Big(\sum\limits_{l=0,1,2}(a_1b_1+\omega^{2l}a_2b_2+\omega^la_3b_3)_{\textbf{1}_l^2}\Big)\\\oplus
\begin{pmatrix}
a_2b_3\\
a_3b_1\\
a_1b_2
\end{pmatrix}_{\textbf{3}_D}\oplus
\begin{pmatrix}
a_3b_2\\
a_1b_3\\
a_2b_1
\end{pmatrix}_{\overline{\textbf{3}}_D}
\end{split}
\end{equation}\\

\begin{equation}
\begin{pmatrix}
a_1\\
a_2\\
a_3
\end{pmatrix}_{\textbf{3}_A}\otimes
\begin{pmatrix}
b_1\\
b_2\\
b_3
\end{pmatrix}_{\textbf{3}_C}=
\begin{pmatrix}
a_1b_1\\
a_2b_2\\
a_3b_3
\end{pmatrix}_{\overline{\textbf{3}}_B}\oplus
\begin{pmatrix}
a_2b_3\\
a_3b_1\\
a_1b_2
\end{pmatrix}_{\overline{\textbf{3}}_C}\oplus
\begin{pmatrix}
a_3b_2\\
a_1b_3\\
a_2b_1
\end{pmatrix}_{\overline{\textbf{3}}_C}
\end{equation}\\

\begin{equation}
\begin{split}
\begin{pmatrix}
a_1\\
a_2\\
a_3
\end{pmatrix}_{\overline{\textbf{3}}_A}\otimes
\begin{pmatrix}
b_1\\
b_2\\
b_3
\end{pmatrix}_{{\textbf{3}}_C}=
\Big(\sum\limits_{l=0,1,2}(a_1b_1+\omega^{2l}a_2b_2+\omega^la_3b_3)_{\textbf{1}_l^2}\Big)\\\oplus
\begin{pmatrix}
a_3b_1\\
a_1b_2\\
a_2b_3
\end{pmatrix}_{\textbf{3}_D}\oplus
\begin{pmatrix}
a_3b_2\\
a_1b_3\\
a_2b_1
\end{pmatrix}_{\overline{\textbf{3}}_D}
\end{split}
\end{equation}\\

\begin{equation}
\begin{split}
\begin{pmatrix}
a_1\\
a_2\\
a_3
\end{pmatrix}_{\textbf{3}_A}\otimes
\begin{pmatrix}
b_1\\
b_2\\
b_3
\end{pmatrix}_{\textbf{3}_D}=
\begin{pmatrix}
a_3b_3\\
a_1b_1\\
a_2b_2
\end{pmatrix}_{\textbf{3}_A}\oplus
\begin{pmatrix}
a_3b_1\\
a_1b_2\\
a_2b_3
\end{pmatrix}_{\textbf{3}_B}\oplus
\begin{pmatrix}
a_3b_2\\
a_1b_3\\
a_2b_1
\end{pmatrix}_{\textbf{3}_C}
\end{split}
\end{equation}\\

\begin{equation}
\begin{split}
\begin{pmatrix}
a_1\\
a_2\\
a_3
\end{pmatrix}_{\overline{\textbf{3}}_A}\otimes
\begin{pmatrix}
b_1\\
b_2\\
b_3
\end{pmatrix}_{\textbf{3}_D}=
\begin{pmatrix}
a_2b_1\\
a_3b_2\\
a_1b_3
\end{pmatrix}_{\overline{\textbf{3}}_A}\oplus
\begin{pmatrix}
a_2b_2\\
a_3b_3\\
a_1b_1
\end{pmatrix}_{\overline{\textbf{3}}_B}\oplus
\begin{pmatrix}
a_2b_3\\
a_3b_1\\
a_1b_2
\end{pmatrix}_{\overline{\textbf{3}}_C}
\end{split}
\end{equation},\\

\begin{equation}
\begin{split}
\begin{pmatrix}
a_1\\
a_2\\
a_3
\end{pmatrix}_{\textbf{3}_B}\otimes
\begin{pmatrix}
b_1\\
b_2\\
b_3
\end{pmatrix}_{\textbf{3}_B}=
\begin{pmatrix}
a_1b_1\\
a_2b_2\\
a_3b_3
\end{pmatrix}_{\overline{\textbf{3}}_B}\oplus
\begin{pmatrix}
a_3b_2\\
a_2b_1\\
a_1b_2
\end{pmatrix}_{\overline{\textbf{3}}_C}\oplus
\begin{pmatrix}
a_2b_3\\
a_1b_2\\
a_2b_1
\end{pmatrix}_{\overline{\textbf{3}}_C}
\end{split}
\end{equation}\\

\begin{equation}
\begin{split}
\begin{pmatrix}
a_1\\
a_2\\
a_3
\end{pmatrix}_{\textbf{3}_B}\otimes
\begin{pmatrix}
b_1\\
b_2\\
b_3
\end{pmatrix}_{\overline{\textbf{3}}_B}=
\Big(\sum\limits_{l=0,1,2}(a_1b_1+\omega^{2l}a_2b_2+\omega^la_3b_3)_{\textbf{1}_l^0}\Big)\\\oplus
\begin{pmatrix}
a_2b_1\\
a_3b_2\\
a_1b_3
\end{pmatrix}_{\textbf{3}_D}\oplus
\begin{pmatrix}
a_1b_2\\
a_2b_3\\
a_3b_1
\end{pmatrix}_{\overline{\textbf{3}}_D}
\end{split}
\end{equation}\\

\begin{equation}
\begin{split}
\begin{pmatrix}
a_1\\
a_2\\
a_3
\end{pmatrix}_{\textbf{3}_B}\otimes
\begin{pmatrix}
b_1\\
b_2\\
b_3
\end{pmatrix}_{\textbf{3}_C}=
\begin{pmatrix}
a_1b_1\\
a_2b_2\\
a_3b_3
\end{pmatrix}_{\overline{\textbf{3}}_A}\oplus
\begin{pmatrix}
a_2b_3\\
a_3b_1\\
a_1b_2
\end{pmatrix}_{\overline{\textbf{3}}_B}\oplus
\begin{pmatrix}
a_3b_2\\
a_1b_3\\
a_2b_1
\end{pmatrix}_{\overline{\textbf{3}}_B}
\end{split}
\end{equation}\\

\begin{equation}
\begin{split}
\begin{pmatrix}
a_1\\
a_2\\
a_3
\end{pmatrix}_{\overline{\textbf{3}}_B}\otimes
\begin{pmatrix}
b_1\\
b_2\\
b_3
\end{pmatrix}_{\textbf{3}_C}=
\Big(\sum\limits_{l=0,1,2}(a_1b_1+\omega^{2l}a_2b_2+\omega^la_3b_3)_{\textbf{1}_l^1}\Big)\\\oplus
\begin{pmatrix}
a_2b_3\\
a_3b_1\\
a_1b_2
\end{pmatrix}_{\textbf{3}_D}\oplus
\begin{pmatrix}
a_1b_3\\
a_2b_1\\
a_3b_2
\end{pmatrix}_{\overline{\textbf{3}}_D}
\end{split}
\end{equation}\\

\begin{equation}
\begin{split}
\begin{pmatrix}
a_1\\
a_2\\
a_3
\end{pmatrix}_{\textbf{3}_B}\otimes
\begin{pmatrix}
b_1\\
b_2\\
b_3
\end{pmatrix}_{\textbf{3}_D}=
\begin{pmatrix}
a_3b_1\\
a_1b_2\\
a_2b_3
\end{pmatrix}_{\textbf{3}_A}\oplus
\begin{pmatrix}
a_3b_3\\
a_1b_1\\
a_2b_2
\end{pmatrix}_{\textbf{3}_B}\oplus
\begin{pmatrix}
a_3b_2\\
a_1b_3\\
a_2b_1
\end{pmatrix}_{\textbf{3}_C}
\end{split}
\end{equation}\\

\begin{equation}
\begin{split}
\begin{pmatrix}
a_1\\
a_2\\
a_3
\end{pmatrix}_{\overline{\textbf{3}}_B}\otimes
\begin{pmatrix}
b_1\\
b_2\\
b_3
\end{pmatrix}_{\textbf{3}_D}=
\begin{pmatrix}
a_2b_3\\
a_3b_1\\
a_1b_2
\end{pmatrix}_{\overline{\textbf{3}}_A}\oplus
\begin{pmatrix}
a_2b_1\\
a_3b_2\\
a_1b_3
\end{pmatrix}_{\overline{\textbf{3}}_B}\oplus
\begin{pmatrix}
a_2b_2\\
a_3b_3\\
a_1b_1
\end{pmatrix}_{\overline{\textbf{3}}_C}
\end{split}
\end{equation}\\

\begin{equation}
\begin{split}
\begin{pmatrix}
a_1\\
a_2\\
a_3
\end{pmatrix}_{\textbf{3}_C}\otimes
\begin{pmatrix}
b_1\\
b_2\\
b_3
\end{pmatrix}_{\textbf{3}_C}=
\begin{pmatrix}
a_1b_1\\
a_2b_2\\
a_3b_3
\end{pmatrix}_{\overline{\textbf{3}}_C}\oplus
\begin{pmatrix}
a_2b_3\\
a_3b_1\\
a_1b_2
\end{pmatrix}_{\overline{\textbf{3}}_A}\oplus
\begin{pmatrix}
a_3b_2\\
a_1b_3\\
a_2b_1
\end{pmatrix}_{\overline{\textbf{3}}_A}
\end{split}
\end{equation}\\

\begin{equation}
\begin{split}
\begin{pmatrix}
a_1\\
a_2\\
a_3
\end{pmatrix}_{\textbf{3}_C}\otimes
\begin{pmatrix}
b_1\\
b_2\\
b_3
\end{pmatrix}_{\overline{\textbf{3}}_C}=
\Big(\sum\limits_{l=0,1,2}(a_1b_1+\omega^{2l}a_2b_2+\omega^la_3b_3)_{\textbf{1}_l^0}\Big)\\\oplus
\begin{pmatrix}
a_2b_1\\
a_3b_2\\
a_1b_3
\end{pmatrix}_{\textbf{3}_D}\oplus
\begin{pmatrix}
a_1b_2\\
a_2b_3\\
a_3b_1
\end{pmatrix}_{\overline{\textbf{3}}_D}
\end{split}
\end{equation}\\

\begin{equation}
\begin{pmatrix}
a_1\\
a_2\\
a_3
\end{pmatrix}_{\textbf{3}_C}\otimes
\begin{pmatrix}
b_1\\
b_2\\
b_3
\end{pmatrix}_{\textbf{3}_D}=
\begin{pmatrix}
a_3b_2\\
a_1b_3\\
a_2b_1
\end{pmatrix}_{\textbf{3}_A}\oplus
\begin{pmatrix}
a_3b_1\\
a_1b_2\\
a_2b_3
\end{pmatrix}_{\textbf{3}_B}\oplus
\begin{pmatrix}
a_3b_3\\
a_1b_1\\
a_2b_2
\end{pmatrix}_{\textbf{3}_C}
\end{equation}\\

\begin{equation} \label{3cb3d}
\begin{pmatrix}
a_1\\
a_2\\
a_3
\end{pmatrix}_{\overline{\textbf{3}}_C}\otimes
\begin{pmatrix}
b_1\\
b_2\\
b_3
\end{pmatrix}_{\textbf{3}_D}=
\begin{pmatrix}
a_2b_2\\
a_3b_3\\
a_1b_1
\end{pmatrix}_{\overline{\textbf{3}}_A}\oplus
\begin{pmatrix}
a_2b_3\\
a_3b_1\\
a_1b_2
\end{pmatrix}_{\overline{\textbf{3}}_B}\oplus
\begin{pmatrix}
a_2b_1\\
a_3b_2\\
a_1b_3
\end{pmatrix}_{\overline{\textbf{3}}_C}
\end{equation}\\

\begin{equation} \label{3d3d}
\begin{pmatrix}
a_1\\
a_2\\
a_3
\end{pmatrix}_{\textbf{3}_D}\otimes
\begin{pmatrix}
b_1\\
b_2\\
b_3
\end{pmatrix}_{\textbf{3}_D}=
\begin{pmatrix}
a_1b_1\\
a_2b_2\\
a_3b_3
\end{pmatrix}_{\overline{\textbf{3}}_D}\oplus
\begin{pmatrix}
a_2b_3\\
a_3b_1\\
a_1b_2
\end{pmatrix}_{\overline{\textbf{3}}_D}\oplus
\begin{pmatrix}
a_3b_2\\
a_1b_3\\
a_2b_1
\end{pmatrix}_{\overline{\textbf{3}}_D}
\end{equation}\\

\begin{equation}
\begin{split}
\begin{pmatrix}
a_1\\
a_2\\
a_3
\end{pmatrix}_{\textbf{3}_D}\otimes
\begin{pmatrix}
b_1\\
b_2\\
b_3
\end{pmatrix}_{\overline{\textbf{3}}_D}=
\Big(\sum\limits_{l=0,1,2}(a_1b_1+\omega^{2l}a_2b_2+\omega^la_3b_3)_{\textbf{1}_l^0}\Big)\\\oplus
\Big(\sum\limits_{l=0,1,2}(a_1b_1+\omega^{2l}a_2b_2+\omega^la_3b_3)_{\textbf{1}_l^1}\Big)\\\oplus
\Big(\sum\limits_{l=0,1,2}(a_1b_1+\omega^{2l}a_2
+\omega^la_3b_3)_{\textbf{1}_l^2}\Big)
\end{split}
\end{equation}\\

The tensor products of singlets are given by
\begin{equation}
\textbf{1}^k_l\otimes\textbf{1}^{k'}_{l'}=\textbf{1}^{k+k'(mod3)}_{l+l'(mod3)}
\end{equation}

\section{The scalar potential and vacuum alignment}
\subsection{The $A_4$ model}
This model contains two scalars: an SM singlet and $A_4$ triplet real scalar field $\chi$ and the SM Higgs which is singlet under $A_4$. The most general renormalizable potential invariant under the SM gauge symmetry and $A_4$ flavour symmetry is written as
\be \label{A4-pot}
V = V(\phi) + V(\chi) + V(\phi, \chi) \ee
where
\beqa
V(\phi) &=& \mu^2 \phi^\dagger \phi + \lambda (\phi^\dagger \phi)^2, \nonumber \\
V(\chi) &=& \mu_\chi^2 (\chi \chi)_{\bf 1} + \sigma (\chi \chi \chi)_{\bf 1} + \lambda_1 (\chi \chi)_{\bf 1} (\chi \chi)_{\bf 1} + \lambda_2 (\chi \chi)_{\bf 1^\prime} (\chi \chi)_{\bf 1^{\prime \prime}} + \lambda_3 (\chi \chi)_{\bf 3} (\chi \chi)_{\bf 3}, \nonumber \\
V(\chi,\phi) &=& \kappa (\phi^\dagger \phi)(\chi \chi)_{\bf 1}~. \eeqa
where $(...)_{\bf r}$ denotes the component of the tensor product of the fields inside the bracket that transform as ${\bf r}$-dimensional irreducible representation.

The minimization consditions of the complete potential evaluated at the required minimum $\vev{\chi_1}=\vev{\chi_2}=\vev{\chi_3} \equiv v_\chi$ lead to
\be \label{A4-minimum}
0=\left[\frac{\partial V}{\partial \chi_i}\right]_{\chi_j=v_\chi} = 2v_\chi (\mu_\chi^2 + \kappa v^2 + 3 \sigma v_\chi + 2(3\lambda_1+ 4 \lambda_3)v_\chi^2)~.\ee
The nontrivial vacuum is obtained as
\be \label{vchi}
v_\chi = \frac{-3\sigma \pm \sqrt{9 \sigma^2 - 8(3\lambda_1+ 4 \lambda_3)(\mu_\chi^2 + \kappa v^2)} } {4 (3\lambda_1+ 4 \lambda_3)}~. \ee
The above minima is global for $\lambda_{1,3} > 0$ and $(\mu_\chi^2 + \kappa v^2)<0$. Therefore in this model the desired vacuum alignment can be obtained without any fine-tunning in the potential. It is easy to see that the required vacuum alignment can be global minimum of $V(\chi)$. The SM Higgs $\phi$, being $A_4$ singlet, does not change this result.

\subsection{The $\Sigma(81)$ model}
The model contains several flavon fields: $\psi_A$, $\psi_B$, $\psi_C$ and $\varphi$ transforming as ${\bf 3}_A$, ${\bf 3}_B$, ${\bf 3}_C$ and ${\bf 3}_D$ respectively. The complete flavon potential can be decomposed into the following pieces for simplicity.
\beqa \label{81-pot}
V &=& V(\varphi) + \sum_{i} \left(V(\psi_i) + V(\phi,\psi_i) \right) + \sum_{i\neq j} \left( V(\psi_i, \psi_j) + V(\phi,\psi_i,\psi_j) \right) \nonumber \\
& + & V(\psi_A,\psi_B,\psi_C) + V(\phi,\psi_A,\psi_B,\psi_C)~, \eeqa
where $i,j=A,B,C$. We find that the desired vacuum of $\varphi$ given in Eq. (\ref{81-varphi-vac}) can be a global minima of $V(\varphi)$. Similar results are found for all the other flavons. However some cross-coupling terms in $V(\phi,\psi_i)$, $V(\psi_i, \psi_j)$, $V(\phi,\psi_i,\psi_j)$, $V(\psi_A,\psi_B,\psi_C)$ and $V(\phi,\psi_A,\psi_B,\psi_C)$ destroy the alignment and require special conditions on the parameters of the most general potential. We therefore find that the desired vacuum of all the flavon fields require several unnatural conditions.

One of the well-known solutions to the vacuum alignment problem is to extend the flavour group as discussed in detail in \cite{Holthausen:2011vd}. The flavour group in this case can be extended in such a way that it preserves the flavour structure and leads to some accidental symmetry in the flavon potential which ensures the desired vacuum structures. Investigation of such possibility is however beyond the scope of the studies presented in this paper and should be taken up elsewhere. We provide a less radical solution to this problem. Some of the unwanted cross-coupling can be avoided by imposing additional symmetries on the flavon potential. Consider a $U(1)$ symmetry under which $\psi_A\to e^{i \alpha} \psi_A$, $\psi_B\to e^{i \beta} \psi_B$, $\psi_C\to e^{i \gamma} \psi_C$  and $\phi \to e^{i \delta} \phi$. Alternatively, $U(1)$ can also be replaced by $Z_n$ symmetry with sufficiently large value of $n$. It is straightforward to see that the invariance of $V$ under $U(1)$ implies
\be \label{p1}
V(\phi,\psi_i,\psi_j) = V(\psi_A,\psi_B,\psi_C) = V(\phi,\psi_A,\psi_B,\psi_C) =0 \ee
in Eq. (\ref{81-pot}). The remaining terms can be obtained using the tensor product rules given in the previous Appendix. They are
\beqa \label{p2}
V(\varphi) & = & \mu_\varphi^2 \phi^\dagger \phi + \sum_{a=1}^6 \lambda^\varphi_a  (\phi^\dagger \phi)_a (\phi^\dagger \phi)_a + \sum_{a=1}^6 \kappa^\varphi_a  (\phi \phi)_a (\phi^\dagger \phi^\dagger)_a~, \nonumber \\
V(\psi_i) & = & \mu_i^2 \psi_i^\dagger \psi_i + \sum_{a=1}^4 \lambda^i_a  (\psi_i^\dagger \psi_i)_a (\psi_i^\dagger \psi_i)_a + \sum_{a=1}^4 \kappa^i_a  (\psi_i \psi_i)_a (\psi_i^\dagger \psi_i^\dagger)_a~, \nonumber \\ 
V(\phi,\psi_i) & = & \sum_{a=1}^3 \lambda^{\varphi i}_a  (\psi_i^\dagger \psi_i)_a (\phi^\dagger \phi)_a + \sum_{a=1}^3 \kappa^{\varphi i}_a  (\psi_i \phi)_a (\psi_i^\dagger \phi^\dagger)_a~, \nonumber \\
V(\psi_i,\psi_j) & = & \sum_{a=1}^5 \lambda^{ij}_a  (\psi_i^\dagger \psi_i)_a (\psi_j^\dagger \psi_j)_a + \sum_{a=1}^3 \kappa^{ij}_a  (\psi_i \psi_j)_a (\psi_i^\dagger \psi_j^\dagger)_a~, \eeqa
where $a=1,2,...$ denotes various possible ways to contract the flavon. For example, in the last term of $V(\phi)$, $a=1,..,6$ correspond to six different ways in which $(\phi \phi)$ and $(\phi^\dagger \phi^\dagger)$ can be contracted to get singlets. They arise from the fact that $(\phi \phi)$ transform as three different $\overline{\bf 3}_D$ representations under $\Sigma(81)$. Since $(\phi \phi)$ and $(\phi^\dagger \phi^\dagger)$ are symmetric, it leads to six independent way to make $(\phi \phi)_a (\phi^\dagger \phi^\dagger)_a$ as singlet.

Using Eqs. (\ref{81-pot},\ref{p1},\ref{p2}) and the vacuum structures in Eqs. (\ref{81-varphi-vac},\ref{vacuum-81}), one obtains
\beqa \label{81-cond}
\left[\frac{\partial V}{\partial \varphi_1}\right]_{\rm min.}&=&\left[\frac{\partial V}{\partial \varphi_2}\right]_{\rm min.} = 0~, \nonumber \\
 \left[\frac{\partial V}{\partial \psi_{A1}}\right]_{\rm min.}&=&\left[\frac{\partial V}{\partial \psi_{A2}}\right]_{\rm min.}=0~, \nonumber \\
 \left[\frac{\partial V}{\partial \psi_{B1}}\right]_{\rm min.}&=&\left[\frac{\partial V}{\partial \psi_{B3}}\right]_{\rm min.}= 0 ~, \nonumber \\
 \left[\frac{\partial V}{\partial \psi_{C2}}\right]_{\rm min.}&=&\left[\frac{\partial V}{\partial \psi_{C3}}\right]_{\rm min.}=0~, \eeqa
 and
 \beqa \label{cond}
\left[\frac{\partial V}{\partial \varphi_3}\right]_{\rm min.} & = & 2 v_\varphi (\mu^2_\varphi + c_1 v_{\psi_A}^2 + c_2 v_{\psi_B}^2 + c_3 v_{\psi_C}^2 + c_4 v_\varphi^2), \nonumber \\
\left[\frac{\partial V}{\partial \psi_{A3}}\right]_{\rm min.} & = & 2 v_{\psi_A} (\mu^2_A + c_{A1} v_{\psi_A}^2 + c_{A2} v_{\psi_B}^2 + c_{A3} v_{\psi_C}^2 + c_{1} v_\varphi^2), \nonumber \\
\left[\frac{\partial V}{\partial \psi_{B2}}\right]_{\rm min.} & = & 2 v_{\psi_B} (\mu^2_B + c_{B1} v_{\psi_A}^2 + c_{B2} v_{\psi_B}^2 + c_{B3} v_{\psi_C}^2 + c_{2} v_\varphi^2), \nonumber \\
\left[\frac{\partial V}{\partial \psi_{C1}}\right]_{\rm min.} & = & 2 v_{\psi_C} (\mu^2_C + c_{A3} v_{\psi_A}^2 + c_{B3} v_{\psi_B}^2 + c_{C3} v_{\psi_C}^2 + c_{3} v_\varphi^2), \eeqa

where
\beqa \label{•}
c_1 &=& \kappa^{\varphi A}_1 + \lambda^{\varphi A}_1 +\lambda^{\varphi A}_3 + \lambda^{\varphi A}_3, \nonumber \\
c_2&=& \kappa^{\varphi B}_2 + \lambda^{\varphi B}_1 + \omega \lambda^{\varphi B}_2 -\omega^2 \lambda^{\varphi B}_3 , \nonumber \\
c_3 &=& \kappa^{\varphi C}_3 + \lambda^{\varphi C}_1 - \omega^2 \lambda^{\varphi C}_2 + \omega \lambda^{\varphi C}_3  , \nonumber \\
c_4&=&2 (\kappa^\varphi_1 +\sum_{a=1}^6 \lambda^\varphi_a), \nonumber \\
c_{A1}&=& 2 (\kappa^A_1 +\sum_{a=1}^3 \lambda^A_a), \nonumber \\
c_{A2}&=&\kappa^{A B}_2 + \lambda^{A B}_1 - \omega^2 \lambda^{A B}_2 + \omega \lambda^{AB}_3 , \nonumber \\
c_{A3}&=&\kappa^{A C}_3 + \lambda^{A C}_1 + \omega \lambda^{AC}_2 - \omega^2 \lambda^{A C}_3  , \nonumber \\
c_{B2} &= & 2 (\kappa^B_1 +\sum_{a=1}^3 \lambda^B_a), \nonumber \\
c_{B3} &=& \kappa^{B C}_2 + \lambda^{B C}_1 - \omega^2 \lambda^{BC}_2 + \omega \lambda^{BC}_3  , \nonumber \\
c_{C3}&=&2 (\kappa^C_1 +\sum_{a=1}^3 \lambda^C_a), \eeqa
The four equations in (\ref{cond}) when equated to zero determine the four VEVs. As it can be seen, there are large number of parameters in potential which lead to the desired values of VEVs. Clearly, no fine-tuning is required to achieve required vacuum alignment. It is to be noted that the $U(1)$ symmetry forbids the terms in Yukawa Langrangian, Eq. (\ref{S81-L}). To avoid this problem, one can introduce a scalar singlet for each flavon triplet with opposite $U(1)$ charge. Each flavon triplet in Eq. (\ref{S81-L}) then can be replaced by the a combination of that flavon triplet and its singlet partner. Since the new scalars are singlets under the full flavour group, they do not modify the vacuum alignment conditions, Eqs. (\ref{81-cond},\ref{cond}). These fields only shift the mass term of flavon triplets in a similar way the VEV of Higgs field $\phi$ changes the $\mu_\chi^2$ to $\mu_\chi^2+\kappa v^2$ as shown in the case of $A_4$ model. The SM Higgs also give rise to the same effects without perturbing the vacuum structure when its interactions are included in $V$ given in Eq. (\ref{81-pot}).

\bibliography{references}
\bibliographystyle{apsrev4-1}
\end{document}